\documentclass[twocolumn]{aa}
\usepackage{graphicx}
\usepackage{txfonts}
\newcommand{\simlt}{\lower.5ex\hbox{$\;\buildrel<\over\sim\;$}}

\begin{document}
   \title{Near-infrared observations of galaxies in Pisces-Perseus}

   \subtitle{V. On the origin of bulges.\thanks{Based on observations at the TIRGO, NOT, and VATT telescopes. 
TIRGO (Gornergrat, CH) is operated by IRA-CNR, Arcetri, Firenze. 
NOT (La Palma, Canary Islands) is operated by NOTSA, the Nordic Observatory Scientific Association. 
VATT (Mt. Graham, AZ) is operated by VORG, the Vatican Observatory Research Group.}}

   \author{L. K. Hunt
          \inst{1}
          \and
	  D. Pierini\inst{2}
	  \and
          C. Giovanardi\inst{3}
          }

   \offprints{L. K. Hunt}

   \institute{Istituto di Radioastronomia-Firenze/CNR,
              Largo E. Fermi 5, I-50125 Firenze, Italy\\
              \email{hunt@arcetri.astro.it}
         \and
           Max-Planck-Institut f\"ur extraterrestrische Physik (MPE),
             Giessenbachstrasse, D-85748 Garching, Germany\\
             \email{dpierini@mpe.mpg.de}
	 \and
	     INAF-Osservatorio Astrofisico di Arcetri,
             Largo E. Fermi 5, I-50125 Firenze, Italy\\ 
             \email{giova@arcetri.astro.it}
             }

   \date{Received ..., 2003; accepted ...}

   \abstract{
We investigate the scaling relations of bulge and disk
structural parameters for a sample of 108 disk galaxies.
Structural parameters of individual galaxies are obtained from
two-dimensional bulge/disk decomposition of their H-band surface brightness distributions\thanks{Table 1 is only
available in electronic form at the CDS via anonymous ftp to cdsarc.u-strasbg.fr
(130.79.128.5) or via http://cdsweb.u-strasbg.fr/cgi-bin/qcat?J/A+A/}.
Bulges are modelled with a generalized exponential (S\'ersic) with variable 
integer shape index $n$.
We find that bulge effective scalelengths $r_e^B$ and luminosity $M^B$
increase with increasing $n$, but disk properties are independent of bulge shape.
As Hubble type $T$ increases,
bulges become less luminous and their mean effective surface brightness $<\mu_e^B>$ 
gets fainter; 
disk $<\mu_e^D>$ shows a similar, but much weaker, trend.
When bulge parameters ($<\mu_e^B>$, $r_e^B$, $M^B$) are
compared with disk ones ($<\mu_e^D>$, $r_e^D$, $M^D$), 
{\it they are tightly correlated for $n=1$ bulges}. 
The correlations gradually worsen with increasing $n$ such that 
$n=4$ bulges appear {\it virtually independent} of their disks.
The Kormendy relation, $<\mu_e^B>$ vs. $r_e^B$, is shown to depend on bulge shape $n$;
the two parameters are tightly correlated in $n=4$ bulges ($r=0.8$), 
and increasingly less so as $n$ decreases;
disk $<\mu_e^D>$ and $r_e^D$ are well correlated ($r=0.7$).
%Luminosity ratios $B/D$ are smaller for $n=1$ bulges;
%they significantly increase with decreasing T, but
%with a surprisingly large spread;
%the value of $B/D$ is governed 
%by the bulge luminosity, since disks have the same average luminosity
%independently of their shape index $n$.
Bulge-to-disk size ratios $r_e^B/r_e^D$ are independent of Hubble type,
but smaller for exponential bulges; 
the mean $r_e^B/r_e^D$ for $n=1$ bulges is 4 times smaller than
that for $n=4$, with a spread which is 9 times smaller.
Strongly barred SB galaxies with exponential bulges are more luminous than their
unbarred counterparts. 
Exponential bulges appear to be closely related to their underlying
disks, while bulges with higher $n$ values are less so;
$n=4$ bulges and their disks apparently have no relation. 
We interpret our results as being most consistent with a secular evolutionary scenario,
in which dissipative processes in the disk are responsible for building up the bulges
in most spirals.
   }

\titlerunning{On the origin of bulges}
\maketitle

\section{Introduction}

This is the fifth paper in a series based on near-infrared imaging of
a sample of disk galaxies in the Pisces-Perseus supercluster.
Already in Paper III of the series (Moriondo et al. \cite{moriondo99b}),
we explored the relations between structural parameters of bulges and
disks, but in the context of distance measurements and the consequent
definition of their Fundamental Planes.
For this reason, that paper: {\it i})~made only use of the subset of
galaxies (40) with measured rotation curves, and {\it ii})~postponed to
future work the discussion of the impact of scaling relations of bulges
and disks on their formation and evolution.
In the present paper, we will address this issue.

The traditional picture is that bulges 
   formed well before the disk of a galaxy, in a ``monolithic collapse'',
either as the first rapid stage of the collapse of a galaxy-size density 
perturbation, as in the Milky Way collapse model of Eggen et al. (\cite{eggen62}),
or through the infall of enriched gas from the star-forming halo
(Carney et al. \cite{carney90}).
``Early'' bulge formation is also predicted within the framework
of hierarchical cold-dark matter (CDM) scenarios of structure formation and evolution
(Kauffmann et al. \cite{kauffmann93}; Baugh et al. \cite{baugh96}).
In this case, the bulge originates from the merging of stellar disks
within the merging halos, while the disk that is observed today
is assembled through gas accretion at later epochs.
A more recent variation of this scenario is the growth of galactic bulges
through merging of dense satellites (Aguerri et al. \cite{aguerri01}).
Evidence in favor of ``early'' bulge formation rises from studies
of the stellar populations in the Milky-Way bulge (Ortolani et al. \cite{ortolani95})
and in bulges of nearby galaxies (Jablonka et al. \cite{jablonka96}).
In addition, bulge velocity fields (Kormendy \& Illingworth \cite{ki82})
and their high central surface brightness seem to be imprints
of a formation process that involved dissipational collapse (e.g., Carlberg \cite{carlberg}) at epochs
earlier than the formation of the disks and, perhaps, of the stellar halo.

More recently, it has been suggested that the bulges in (some) disk galaxies
can be formed after the disk, through secular evolutionary phenomena.
Numerical simulations (Combes \& Sanders \cite{combes81}; Combes et al. \cite{combes90};
Norman et al. \cite{norman96}) have shown that a spheroidal stellar component can originate from
dissipationless processes such as the thickening or destruction of a bar.
Moreover, when gas is included in the simulations, dissipation accelerates the
evolution, giving rise to galaxies which are more centrally concentrated
(Junqueira \& Combes \cite{junqueira}).
Indeed, some of the observed bulge characteristics such as triaxial or peanut shape and cylindrical rotation 
(e.g., Shaw \cite{shaw93}; L\"utticke et al. \cite{lutticke}) are predicted by such models.
However, gas is not necessary for dissipation, as pointed out by Zhang in a series of
papers (Zhang \cite{zhang96}, \cite{zhang98}, \cite{zhang99}).
Secular evolution may also operate through outward transport of 
angular momentum through a collective dissipative process linked to spiral
density waves.
Observational evidence in favor of the secular scenario include the similarity of bulge 
and disk color indices (e.g., Balcells \& Peletier \cite{balcells94}), 
the mass-to-luminosity ratios of the components (Widrow et al. \cite{widrow}),
the disk-like kinematics
observed in  many bulges (Kormendy \cite{kormendy93}), and the correlation of bulge and disk scalelengths
(de Jong \cite{dejong96}).
Indeed, when a simple exponential is used to fit the bulge surface brightness profile,
a strong correlation is found between bulge and disk scalelengths
(Courteau et al. \cite{courteau96}), seemingly supporting the secular formation scenario for bulges.
The correlation was later shown to depend critically on the use of a simple exponential
as a bulge fitting function, since when generalized exponentials were used
instead, the mean scalelength ratio depends on the bulge shape index
(Moriondo et al. \cite{moriondo98a} -- hereafter MGH; Graham \cite{graham01}).
Thus it appears, albeit for small samples, that the ratio of bulge and disk scalelengths
may not be constant over a range of Hubble types as previously claimed.
This is of importance not only for the chronology of bulge and disk formation,
but also for the meaning of the Hubble sequence itself in terms of photometric
observables (scale-free Hubble sequence).

To further investigate bulge and disk structural parameters, and to better constrain
their connection with formation scenarios, we have performed two-dimensional
(2D) bulge/disk decompositions on a large sample of disk galaxies in
the Pisces Perseus supercluster (see Moriondo et al. \cite{moriondo99a}, hereafter Paper I).
These decompositions are performed on image data in the H band
because the near-infrared (NIR) wavelengths are the most direct tracer
of the stellar mass (e.g., Rix \& Rieke \cite{rix}, Gavazzi \cite{gavazzi93}).
Moreover, NIR emission is much less affected by 
recent episodes of star formation which typically dominate 
disk morphology at optical wavelengths.  
Finally, the effects of dust extinction are mitigated at 1.65$\,\mu$m (H) 
since $A_H$ is roughly a factor of 8 lower than $A_V$.
We describe the sample and the imaging data in Sect. \ref{sample},
together with the bulge/disk decompositions. 
Section \ref{results} presents our results, and we interpret them
in Sect. \ref{discussion} in the context of various evolutionary scenarios.

\section{The sample, the images, and the structural decomposition \label{sample}}

The galaxies under investigation belong to a sample
selected from the catalogued disk galaxies in the Pisces-Perseus (PP)
supercluster area, $\rm 22^h < RA < 4^h$ and $\rm 0^\circ < \delta < 45^\circ$,
excluding obvious foreground members with heliocentric velocities
less than $\rm 3000~km~s^{-1}$.
The original sample was restricted to about 950 galaxies
with optical major axis $0^{\prime}.5 < a < 4^{\prime}.0$
and available good-quality 21-cm spectra (Giovanelli \& Haynes \cite{giovanelli89}
and references therein).
From this set two distinct subsamples were extracted.
Sample A comprises about 150 galaxies, randomly selected to cover
the entire area and inclination range, and with types Sb or later.
Sample B comprises the 50 cluster galaxies within an angular distance
of $\rm 5.5^\circ$ from the Pisces cluster center ($\rm 1^h~20^m$,~$\rm 33^\circ$),
with optical size $a \ge 0^{\prime}.8$ and inclination
$\rm 30^\circ <i< \rm 75^\circ$, spanning the range S0 to Sd in Hubble type.

Out of these 200 objects, images were obtained for 174 galaxies (see Paper I) 
in the H band (35 in JHK) with the ARNICA camera (Lisi et al. \cite{lisi93},
\cite{lisi96}; Hunt et al. \cite{hunt96}) mounted on various telescopes.
We refer the reader to Paper I for details concerning
observations, data reduction, and photometric calibration.
Analysis of the JHK color images is presented in Moriondo et al. (\cite{moriondo01}, Paper IV).
Unless otherwise specified, all the photometric data derived and here reported refer to the 
set of images in the H band.

\subsection{Two-dimensional bulge/disk decomposition \label{fits}}

%-------------------------------------------------------------
\begin{figure}
\resizebox{9cm}{!}{\includegraphics*{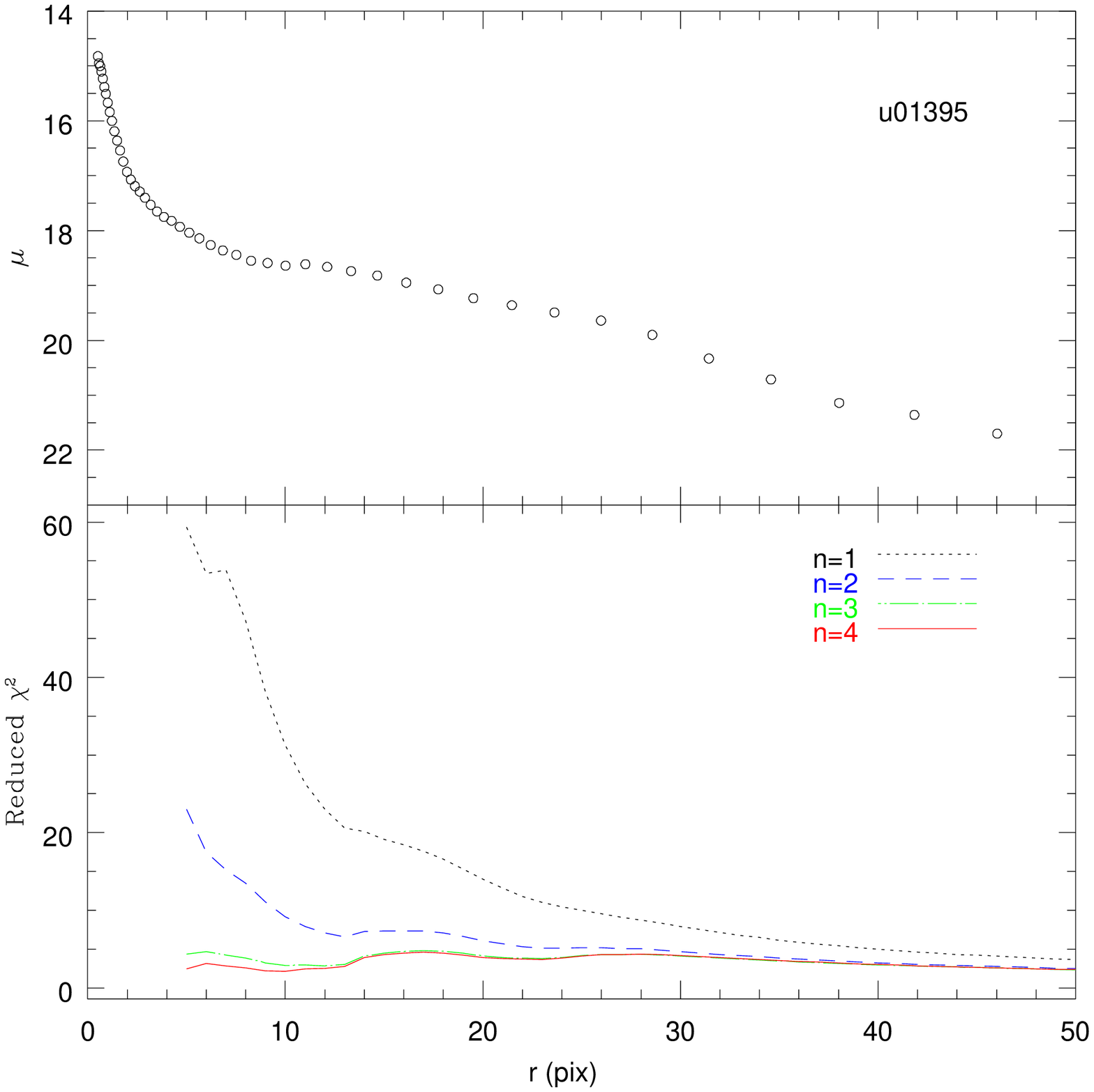}}
\resizebox{9cm}{!}{\includegraphics*{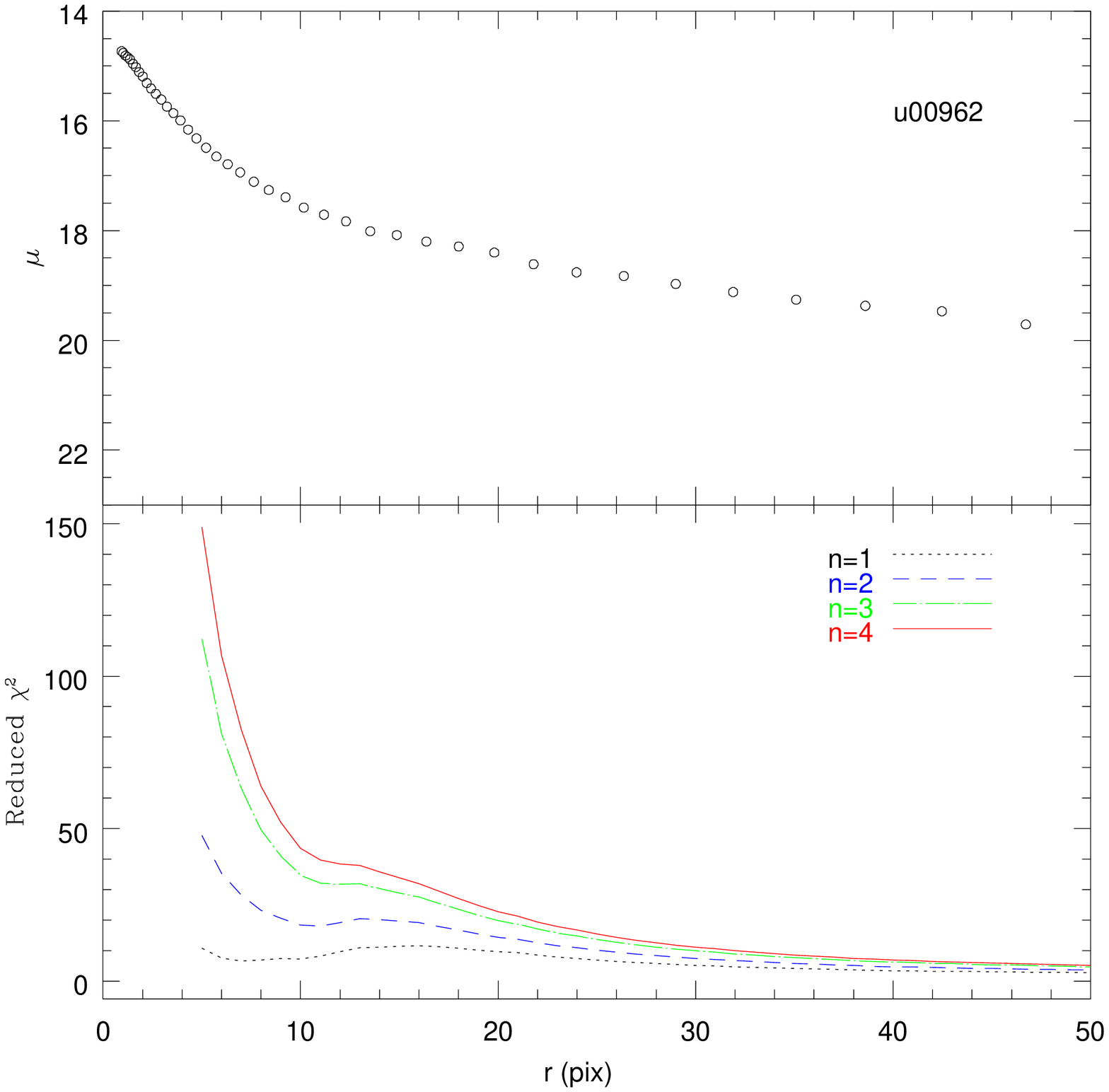}}
\caption{Upper panels: surface brightness profile along the major axis (from elliptical isophotal
fitting) in H mag arcsec$^{-2}$. Lower panels: radial run of the reduced
$\chi^2$ of the best-fitting 2D model for different values of the $n$.
The first panel shows a galaxy with a best-fit $n$ of 4, while a best-fit $n=1$ is shown
in the second panel. Abscissas are labelled in pixel, 1 pixel\,=\,0.975 arcsec.}
\label{fig:chisq}
\end{figure}
%
%-------------------------------------------------------------
As in MGH, we fitted the images with a two-dimensional (2D) parametric brightness distribution model,
consisting of a generalized exponential (S\`ersic 1968) bulge of shape index $n$ :

\begin{equation}
I^B\,(x,y)\ =\ I_e\exp
\left\{-\alpha_n \left[\left(\frac{1}{r_e}\sqrt{x^2+\frac{y^2}
{(1-\epsilon^B)^2}}\ \right)^{1/n}-1\right]\right\}\;\;,
%\label{bulge_b}\protect
\end{equation}

\noindent 
and an exponential thin disk:

\begin{equation}
I^D\,(x,y)\ =\ I_d(0)\exp\left[-\frac{1}{r_d}\sqrt{x^2+
\frac{y^2}{\cos^2 \,i}}\ \right] \; .
%\label{disk_b}\protect
\end{equation}

\noindent
$I_e$ and $r_e$ are effective (half-light) values,
$\epsilon^B$ is the apparent bulge ellipticity,
$\alpha_n$ is a constant relating
the effective brightness and radius to the exponential values (see MGH).
$x$ and $y$ are in arbitrary units, with origin
at the galaxy's center, and
$x$ along the major axis.
The bulge is assumed to be an oblate rotational ellipsoid
(Kent et al. \cite{kent91}), coaxial with the disk, and its
apparent eccentricity $e^B$ is related to
the intrinsic eccentricity $e'^B$ by:
\begin{equation}
\label{ecce}\protect
e^B = e'^B\sin i \;\; .
\end{equation}

The structural parameters have been determined
by fitting the 2D image to the model, convolved with a circular
Gaussian seeing disk of appropriate FWHM, using a $\chi^{2}$ minimization.
The parameters left free to vary in the fit include:
the two surface brightnesses $I_e$ and $I_d(0)$ (or
$\mu_e$ and $\mu_0$ when given in magnitudes),
the two scale lengths $r_e$ and $r_d$, the bulge ellipticity
$\epsilon^B$, and the system inclination $i$.
The fitting algorithm and tests of its reliability are described
more fully in MGH.

The use of a 2D fitting technique is mandatory in the analysis of objects
of middle to high inclination, such as samples -- including the present one 
(68 galaxies with $i>40^\circ$) -- selected for Tully-Fisher studies.
A feature of our algorithm is its ability to suppress non-axisymmetric structure.
This is achieved through a ``folding'' procedure in 
which we exploit the model symmetry about the major and minor axes.
The galaxy image is first ``folded'' about the major axis (measured
a priori), in order to obtain the ``long'' semi-ellipse;
it is then folded again about the minor axis in order to obtain only
one quadrant. 
%which the four quadrants are folded around the major and minor axis.
There are consequently four points used to infer the value of a single pixel 
in the folded quadrant, and from these, a
mean and a standard deviation are calculated; only this single quadrant is input to
the fitting routine.
Typically pixels are weighted by the photon statistics, calculated
by modelling the detector noise characteristics and assuming
background limited performance (verified to be true in all cases). 
The weighting scheme however gives greater weight to those regions with low standard deviation
since it chooses the maximum of the Poisson statistics or the standard deviation
resulting from the folding procedure.
The regions with significant asymmetry show up 
with large standard deviation in the folding process,
and they are given accordingly less weight in the decomposition.

Because of the known mutual functional dependency of $r_e$, $\mu_e$, and $n$,
we chose to use a "grid method" for the determination of $n$ rather than include it as a free
parameter (see also MacArthur et al. \cite{macarthur03}, hereafter MACH03).
Because of the large sample and consequent numerical calculation time for the 
2D fits, we restricted n to integer values of 1, 2, 3 or 4, and ran fits for each of
these separately.
The best-fitting $n$ value was then derived by minimizing the reduced $\chi^2$ over the
inner region.
Two examples of the radial run of $\chi^2_\nu$ for different $n$'s
are shown in Fig. \ref{fig:chisq}.
The surface brightness profiles in the upper panels 
are derived from elliptical isophotal fitting and are only used for display purposes;
the lower panels show the behavior of the reduced $\chi^2$ resulting from the
2D best fit at different $n$, averaged within the same elliptical
annuli of the upper panel.
As can be seen,
the bulge fit influences primarily the inner part of the galaxy.
It is consequently necessary to consider the value of $\chi^2_\nu$ as a function of $n$
especially at small radii, because the large area of the disk tends to make the influence of 
the bulge shape in the global $\chi^2$ very small and easily masked by other features.
The median (and mean) FWHM seeing of the images in our database is 1\farcs9.
Only 4 of 108 galaxies have $r_e^B$ smaller than this, but it
is difficult to distinguish star clusters or nuclear point sources from 
high-n bulges. Point sources have not been included in our fits.

The discretized range of integer $n$ values
1, 2, 3, 4 samples rather well the continuous range of $n$ values
found by previous authors.
MACH03, on the basis of one-dimensional (1D) profile fits, conclude that the 
surface brightness profiles of Freeman Type-I galaxies (Freeman \cite{freeman70}) 
are adequately described by a double-exponential distribution with
bulges of $0.1 < n < 2$;
however only 8 of the 121 galaxies studied by them have $n < 0.5$.
On the other hand, with 2D fits,
M\"ollenhoff \& Heidt (\cite{moellenhoff02}, hereafter MH01) 
found bulge shape indices of $0.8<n<6.7$
for their sample of 40 bright galaxies, with
only one object having a bulge with $n<1$ and two objects with $n>4$.

The fits were judged to be reliable ($\chi^2_\nu \leq 2$) in 117 (of 174) galaxies.
The average percentage errors for the fitted parameters of bulge and disk
are less than 20\%, consistent with MH01 and MACH03.
We have further restricted the final database to contain only those galaxies
with apparent optical diameter $a \geq 1 \arcmin$ 
(from NED\footnote{The NASA/IPAC Extragalactic Database (NED) is operated by the 
Jet Propulsion Laboratory, California Institute of Technology, under contract 
with the National Aeronautics and Space Administration.}). 
Thus, the final PP sample consists of 108 galaxies, 2 of which do not have
well-defined morphological types, and to which we have assigned T=5 (following
LEDA, the Extragalactic Database based in Lyon).   
The best fit parameters resulting from our 2D fits for these 108 disk galaxies
are reported in Table 1.
   Distances to individual galaxies are determined from the Hubble law
   (with $H_0=70 \rm ~km~s^{-1}~Mpc^{-1}$) after correcting
   the heliocentric radial velocities listed in Paper I
   to the CMB reference frame according to the prescription of 
the Third Reference Catalogue (de Vaucouleurs et al. \cite{devaucouleurs91}, hereafter RC3). 
The full table is only available in electronic form, please see the footnote to the Abstract,
here we reproduce it partly to illustrate its contents.
Table 1 also contains additional information from RC3
as follows:\newline 
   {\it Column \ \ \ \ ~~~~~1}: UGC galaxy number, if not differently specified; \newline
   {\it Column \ \ \ \ ~~~~~2}: Distance (Mpc); \newline
   {\it Column \ \ \ \ ~~~~~3}: Hubble type $T$; \newline
   {\it Column \ \ \ \ ~~~~~4}: bar class ;\newline
   {\it Column \ \ \ \ ~~~~~5}: global $\chi^2_\nu$ of fit;\newline
   {\it Column \ \ \ \ ~~~~~6}: bulge shape index $n$ ($n=0$ for pure disks);  \newline
   {\it Column \ \ \ \ ~~~~~7}: apparent bulge ellipticity $\epsilon^B$; \newline
   {\it Columns \ ~~~8-9}: bulge effective radius $r_e$ and uncertainty (arcsec); \newline
   {\it Columns 10-11}: bulge effective surface brightness, reduced to face-on, 
   $\mu_e$ and uncertainty ($\rm mag~arcsec^{-2}$); \newline
   {\it Column \ \ \ ~~~~12}: total bulge apparent magnitude $m^B$ (mag); \newline
   {\it Column \ \ \ ~~~~13}: disk inclination $i$ (degrees);  \newline
   {\it Columns 14-15}: disk scalelength $r_d$ and uncertainty (arcsec); \newline
   {\it Columns 16-17}: central face-on disk surface brightness $\mu_0$
   (and uncertainty $\rm mag~arcsec^{-2}$);\newline
   {\it Column \ \ \ ~~~~18}: total disk apparent magnitude $m^D$ (mag); \newline
   {\it Column \ \ \ ~~~~19}: bulge-to-disk luminosity ratio $B/D$; \newline
   {\it Column \ \ \ ~~~~20}: total galaxy apparent magnitude $m$ (mag).

\setcounter{figure}{1}

\subsection{Photometric corrections}

The magnitudes have been corrected for Galactic extinction according to the estimates
given by Burstein \& Heiles (\cite{burstein}) as reported by NED, and assuming the
extinction curve of Cardelli et al. (\cite{cardelli}).
Using the recipe of Frogel et al. (\cite{frogel}),
we have applied corrections for redshift dimming.

While internal dust extinction may affect the estimates of structural parameters even in the
NIR (e.g., Peletier \& Willner \cite{peletier}),
the corrections are small and subject to some uncertainty.
An empirical analysis
of the photometric corrections for dust extinction
in this sample has been carried out by Moriondo et al. (\cite{moriondo98b}, hereafter Paper II).
To do this, they rely on
correlations with inclinations of
the observed disk and bulge structural parameters and colors, and find
that
the structural parameters of the bulge are not affected by dust extinction,
while those of the disk may be -- especially the disk scalelength -- but
with large scatter. 
However, the analysis Paper II is based on the assumption
that dust is homogeneously distributed
within a thin disk coincident with the mid-plane of the stellar distribution,
which in general is not true
(e.g. Bianchi et al. \cite{bianchi96}, \cite{bianchi00}; Ferrara et al. \cite{ferrara99}).
Moreover,
the behavior of the kinematic scalelength as a function of
the axial ratio is complex (Giovanelli \& Haynes \cite{giovanelli02}, see also Paper II).
Therefore, for the subsequent analysis, the structural parameters
($r_e$, $\mu_e$, $m^B$, $r_d$, $\mu_0$, $m^D$, $B/D$ and $m$)
are corrected to face-on assuming full transparency, which implies no
corrections to scalelengths, total magnitudes, and bulge-to-disk
luminosity ratio $B/D$.

All the above corrections have already been applied to the data reported in Table 1.

%The use of parameters not corrected for dust extinction
%also allows us to compare our results with previous studies
%(e.g., MH01; MACH03).

\subsection{Comparison with previous work}

Our sample contains five galaxies (UGC\,463, UGC\,927, UGC\,1089, UGC\,12378, UGC\,12527) 
in common with MACH03 (H band).
In all cases, the two sets of disk parameters agree quite well:
$r_d$(ours)/$r_d$(theirs)\,=\,0.95\,$\pm$\,0.05 and
$\mu_0$(ours)-$\mu_0$(theirs)\,=\,0.16\,$\pm$\,0.06 H-mag\,arcsec$^{-2}$.
For UGC\,927, they classify their profile as a ``truncated disk'', but it appears very different
from our major-axis cut (all our fits can be obtained in graphic form by request from the authors).
They find $n=2.7$, the largest in their sample, while we find $n=4$;
the bulge surface brightnesses differ by about $\sim$\,1~mag.
This may be an example of the pitfalls of 1D fitting since this galaxy has twisted isophotes,
and marked spiral structure even in the NIR.
Such morphological features tend to be suppressed by our fitting algorithm, and
the greater number of degrees of freedom in 2D image fitting is clearly an advantage.

In general, we find larger $n$ values than those by MACH03: the mean difference
$\Delta n\,=\,1.1\,\pm\,0.75$. 
We also find larger bulge $r_e$, but this is expected because of the different $n$'s.
This may be another advantage of our fitting algorithm, as it takes into account
the non-spherical shape of the bulge (e.g., MGH).
Bulges are rarely spherical (Kent \cite{kent88}), and not considering
the possible bulge ellipticity would result in a steeper bulge outer profile, and thus
a lower $n$.

\section{Results \label{results}}
    
A priori the parent sample suffers from a known dearth of early-type spirals
(e.g., $T\leq 2$, see Paper I). 
Indeed our present sample contains only 6 galaxies with type Sab or earlier, with a
sample median of $T\,=\,5$ (Sc). 
Therefore, we have added to the subsequent analysis the sample of 14 early-type spirals (Sa's)
studied by MGH;
the same fitting algorithm was used to decompose the images.\footnote{A 
correction of 0.25~mag has been made to convert the K-band quantities in that paper 
to the H band used here (e.g., Giovanardi \& Hunt \cite{giova96}).
The distance scale has also been transformed to the $H_0$ (70~km\,s$^{-1}$\,Mpc$^{-1}$)
adopted for this study.
}
The combined sample comprises 122 galaxies, now with 21 of type Sab or earlier, and with median
T\,=\,4 (Sbc). 

Throughout the analysis, we will make use of mean
effective surface brightnesses (SFBs), calculated as the average flux within a circular area of radius $r_e$.
The disk scalelengths $r_d$ are converted to effective ones $r_e^D$ by multiplying by
a factor of 1.6784 appropriate for $n=1$ (see MGH),
and we define the mean effective SFB $<\mu_e^B>$ and $<\mu_e^D>$ for bulge and disk respectively.
In the following, to avoid confusion, the bulge scalelength will be indicated as $r_e^B$.
Mean SFBs $<\mu_e>$ were chosen, rather than the central $\mu_0$ 
(Khosroshahi et al. \cite{khosroshahi00a}, \cite{khosroshahi00b}),
or $\mu_e$ -- the SFB at $r_e$ (e.g., Graham \cite{graham01}; MACH03) -- 
because of the strong functional dependence on $n$ of the latter parameters.
If a given bulge is fit with a generalized exponential, $\mu_e$ becomes fainter
as $n$ of the fit becomes larger, because of the mathematical form of the generalized
exponential. 
Higher $n$ fits are more strongly peaked and consequently have a higher
$\mu_0$, but are more extended, hence a larger $r_e$, 
a fainter $\mu_e$, and a higher bulge luminosity $M^B$
(cf., Fig. 2 in MGH).
In practice, when the same bulge is fit with $n$ varying between 1 and 4, the resulting $\mu_e$ changes
typically by more than 3~mag\,arcsec$^{-2}$, while $<\mu_e>$ also becomes fainter but by $\simlt\,$1~mag\,arcsec$^{-2}$.
The adoption of $<\mu_e>$ also conforms to the definitions of RC2 and RC3, 
and to all the lore of the Fundamental Plane (FP). 

\subsection{Pure disk galaxies}

Only two of the galaxies (UGC\,975 and UGC\,1579) are best fit with pure disks ($B/D\,=\,0$);
bulges are found in all the others, although in some cases as faint as 16.9 H mag,
with $B/D$ ratios of 1\%.
UGC\,975 is classified as a generic spiral (``S''), 
as it shows on blue images a tight flocculent pattern
and high surface brightness.
UGC\,1579, an almost face-on SB(s)d, shows an extended spiral structure and a 
large central region of relatively high surface brightness.
They are among the faintest objects in the sample, but still galaxies with $M\sim-22$
H mag, and are not low surface brightness galaxies.
Pure disks are rare in our sample, although it comprises 50 galaxies with T$\geq$5.
The selection criteria adopted tend to avoid very late types of low surface brightness;
for this reason, our two pure disks are unlikely to be representative of their class
(cf. Gavazzi et al. \cite{gavazzi00}).

%-------------------------------------------------------------
\begin{figure}
\centering
\resizebox{\hsize}{!}{\includegraphics[width=12cm]{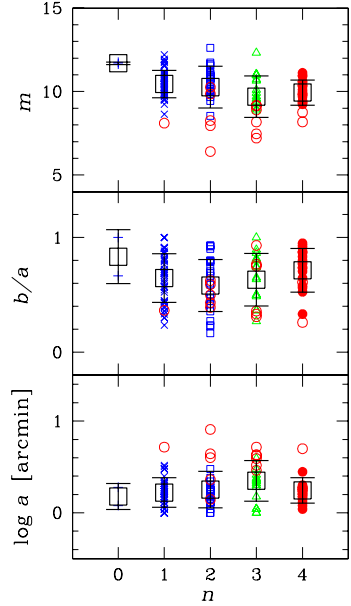}}
\caption{Distributions of H-band apparent magnitude $m$ (top panel),
optical axial ratio $b/a$ (middle), and apparent optical size $a$ (lower) as a function
of bulge shape index $n$. 
Galaxies best fitted with $n=1,2,3,4$ are shown by $\times$, open squares, open triangles, and filled circles,
respectively.
The early-type spirals from MGH are also included and shown as open circles independently
of their $n$.
Data for pure-disk galaxies are shown by $+$ at $n=0$.
Large open squares correspond to mean values, and the error bars to $1\sigma$ standard deviation.
         \label{fig:bias}}
   \end{figure}
%-------------------------------------------------------------
%-------------------------------------------------------------
%
   \begin{figure}
   \centering
\resizebox{\hsize}{!}{\includegraphics[width=12cm]{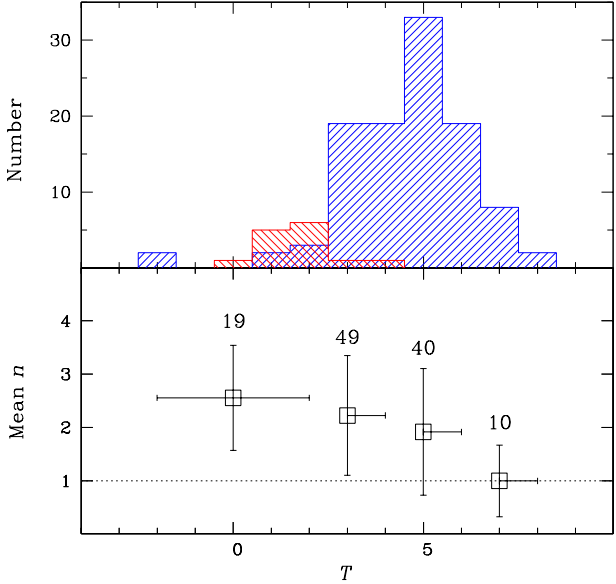}}
      \caption{Distribution of Hubble type $T$ (upper panel) and
bulge shape index $n$ versus $T$ (lower panel).
In both panels, as in all subsequent figures, the galaxies from MGH are 
included. In the histogram, the shading for the PP sample is right slanted
and for the MGH sample is left slanted. 
In the lower panel, horizontal error bars indicate the range
of Hubble type over which the mean $n$ is calculated.
The vertical bars are 1$\sigma$ standard deviation.
The horizontal dotted line is placed at $n=1$ for convenience.
The number of objects in each
$T$ bin is printed over its data point.
              }
         \label{fig:T_n}
   \end{figure}
%---------------------------------------------------------------

\subsection{Possible biases \label{bias}}

In principle, the performance of our decomposition procedure could depend on
the size and quality of the image.
We have performed a simple check for such biases by examining the distributions of
apparent image parameters versus the bulge shape index $n$, as shown in 
Fig. \ref{fig:bias}; 
optical properties are taken from RC3.
The only evidence is that:
({\it i})~the two pure-disk ($n=0$) galaxies are faint, and 
({\it ii})~the Sa's from MGH are substantially brighter than the PP set.
For the rest, no obvious significant trends are detectable.
In particular, we observe no trend for faint, small, and highly inclined galaxies,
that is those observed with lower relative resolution, 
to be better fit with any particular $n$ values.
This is important for the reliability of any assertions regarding trends with 
bulge shape index.

\subsection{Hubble type and bulge shape $n$}

The upper panel of Fig. \ref{fig:T_n} shows the distribution of Hubble types $T$ 
for our combined sample.
The PP and MGH samples have different shadings and
the paucity of early-type spirals in the original PP sample is evident. 

The lower panel shows the means of the bulge shape index $n$ averaged over different
Hubble types. Here and in the following we have grouped Hubble types into
four bins: $T \leq 2$, $2<T\leq4$, $4<T\leq6$, and $T\geq7$.
It can be seen from the figure that $n=1$ bulges tend to be more common in later-type
spirals, and that early types ($T \leq 2$) tend to have bulges with larger $n$,
significantly different from $n=1$.
This result agrees with other authors (Andredakis et al. \cite{andredakis95}; Graham \cite{graham01}) 
who found that $n$ tends to increase with decreasing Hubble type.
Such a trend could also explain the inconsistency with the results of MACH03 
who found no evidence for bulges with $n>2$, but whose sample contains {\it only two galaxies 
earlier than} Sb ($T<3$).

In what follows, we will examine trends of structural parameters both
with the bulge index $n$ and with Hubble type $T$.
The justification for this is that $n$ and $T$ are only weakly correlated
(see Fig.\ref{fig:T_n} and previous work).
We want to investigate trends with {\it bulge shape} as well as with Hubble type
since bulges with different $n$ may be intrinsically different
(e.g., Carollo \cite{carollo99}; Carollo et al. \cite{carollo01}), independently of the Hubble type
of their host galaxies. 
Therefore, with the aim of singling out the more significant parameter,
bulge shape $n$ or Hubble type $T$, we will analyze trends of photometric
parameters with both. 

\subsection{Bulge/disk parameters versus bulge shape $n$ \label{bdn}}

%-------------------------------------------------------------
   \begin{figure}
   \centering
\resizebox{\hsize}{!}{\includegraphics[width=15cm]{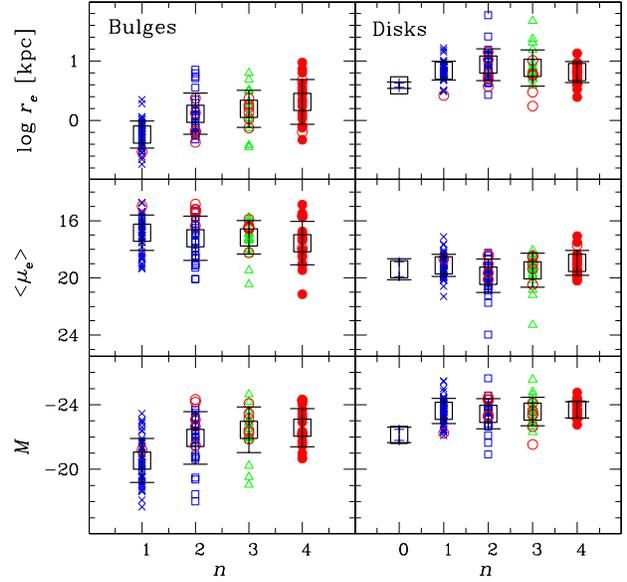}}
      \caption{Bulge and disk structural parameters versus bulge shape index $n$.
The left panels show bulge effective scalelengths (upper panel),
mean effective surface brightnesses (middle), and absolute magnitude (lower).
The right panels show the same quantities for the disks.
Symbols as in Fig. \ref{fig:bias}.
              }
         \label{fig:bd_n}
   \end{figure}
%
%-------------------------------------------------------------
The behavior of the structural parameters of bulges and disks as a function
of the shape index $n$ of the bulge is illustrated in Fig. \ref{fig:bd_n}.
Inspection of the figure shows that
neither bulge nor disk $<\mu_e>$ vary significantly with $n$.
Obviously, when the bulge $\mu_e$ or $\mu_0$ are used (not shown), 
there is a large variation which is
a result of the functional form of the fitting law. 
The increase of central bulge surface brightness $\mu_0$ with $n$
is $\sim\,$6 mag arcsec$^{-2}$ in agreement with 
Khosroshahi et al. (\cite{khosroshahi00a}).

$r_e^B$, instead, increases significantly with $n$: 
the left panel of Fig. \ref{fig:bd_n} shows that $n=1$ bulges have
smaller scalelengths than bulges with higher $n$\footnote{This effect is 
significant at $>$\,99.9\% level (U test).
However, the significance is only obtained for $n=1$ bulges;
the differences among $r_e^B$ at other $n$ values are not significant. }, while
there is no dependence of disk scalelengths on the $n$ of the bulge.
The increase of $r_e^B$ with $n$ is consistent
with earlier work on ellipticals (Caon et al. \cite{caon}), on early-type spirals 
(Khosroshahi et al. \cite{khosroshahi00a}), and on the later-type spirals in the de Jong sample
(de Jong \& van der Kruit \cite{dejong94}; Graham \cite{graham01}). 
Such a trend was shown to be real, rather than a result of the mutual dependence
among $n$, $r_e^B$, and $\mu_e$ of the bulge fitting law, by Trujillo et al. (\cite{trujillo01}).
Moreover, since the properties of $n=1$ bulges are shown to differ 
{\it with high statistical significance} from their higher $n$ counterparts,
it vindicates our method which relies on integer values of $n$.

The lower panels of Fig. \ref{fig:bd_n} shows $M^B$ and $M^D$ as a function of $n$.
In agreement with earlier work 
(Andredakis et al. \cite{andredakis95}; Graham \cite{graham01}; MH01; Trujillo et al. 2002), 
bulges with larger $n$ tend to be more luminous, 
although with some scatter.
Again, we find that $n=1$ bulges are significantly different from bulges 
with higher $n$, being
{\it less luminous} with a significance $>$\,99.9\%. 
On the other hand, $M^D$ appears to be independent of $n$, and significantly
less variable in general:
the standard deviations of $M^B$ range from 1.2 to 1.6 mag, while those of $M^D$
are considerably smaller, 0.5 to 0.8 mag.
%Since absolute magnitude depends exclusively on $r_e$ and $<\mu_e>$, the higher luminosity
%of higher $n$ bulges must be ascribed to the relatively larger effective radii in these.
% this statement is false because of the n factor.
The SFBs do not vary with $n$, and cannot be the cause of
the dependence of $M^B$ on $n$.
Exponential bulges are {\it smaller} and {\it less luminous} than their higher $n$
counterparts, but roughly the same in terms of mean surface brightness.
None of the disk properties ($<\mu_e^D>$, $r_e^D$, $M^D$) 
depends on the $n$ value of the bulge.

\subsection{Bulge/disk parameters versus Hubble type \label{bdt}}

%-------------------------------------------------------------
   \begin{figure}
   \centering
\resizebox{\hsize}{!}{\includegraphics[width=12cm]{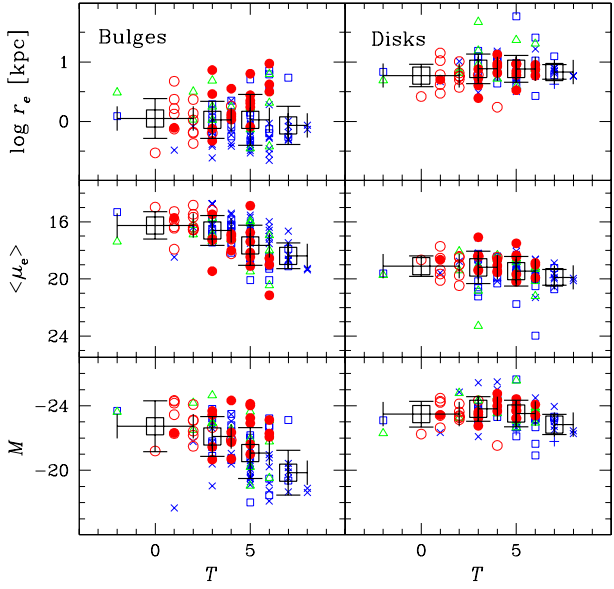}}
      \caption{Bulge and disk structural parameters versus Hubble type $T$.
The left panels show bulge effective scalelengths (upper panel),
mean effective surface brightnesses (middle), and absolute magnitude (lower).
The right panels show the same quantities for the disks.
Symbols as in Fig. \ref{fig:bias}.
              }
         \label{fig:sbmean_t}
   \end{figure}
%
%-------------------------------------------------------------
Here we examine the same parameters of the previous section, but now
consider their variation with Hubble type $T$.
The relevant graphs are presented in Fig. \ref{fig:sbmean_t}.

%-------------------------------------------------------------
   \begin{figure}
   \centering
\resizebox{\hsize}{!}{\includegraphics[width=15cm]{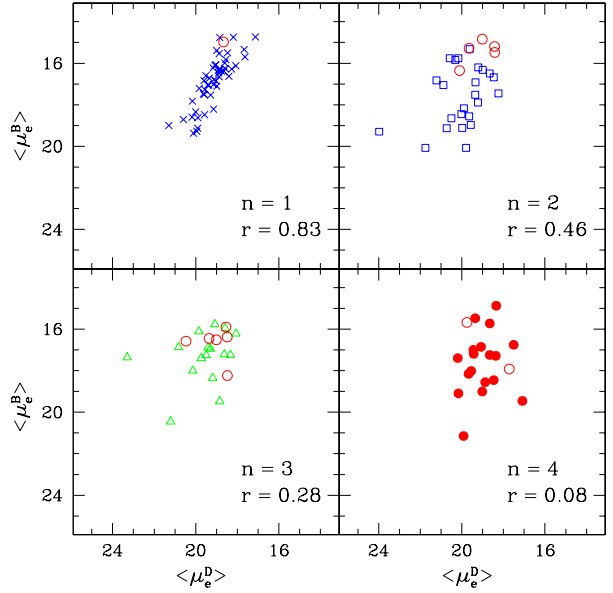}}
      \caption{Bulge versus disk mean surface brightnesses $<\mu_e>$,
separately for galaxies with different best $n$ values of the bulge.
The correlation coefficients are shown below the values of $n$.
              }
         \label{fig:sb_n}
   \end{figure}
%
%-------------------------------------------------------------

%-------------------------------------------------------------
   \begin{figure}
   \centering
\resizebox{\hsize}{!}{\includegraphics[width=15cm]{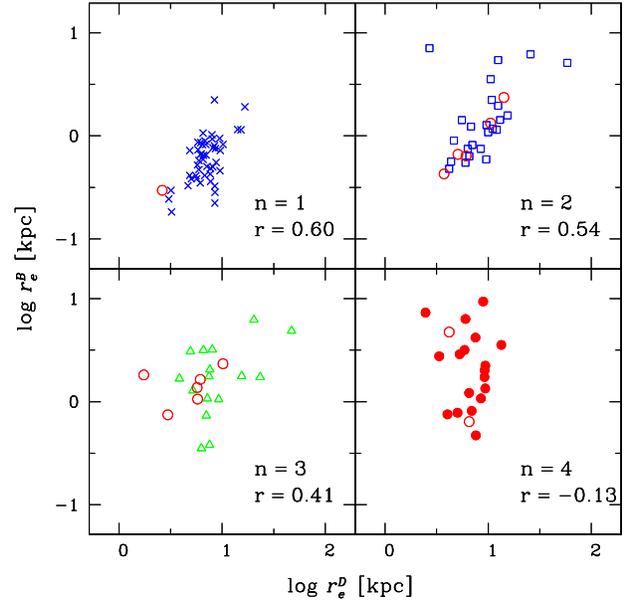}}
      \caption{Bulge versus disk effective scalelengths $r_e^B$, $r_e^D$,
separately for galaxies with different best $n$ values of the bulge.
The correlation coefficients are shown below the values of $n$.
              }
         \label{fig:logre_n}
   \end{figure}
%
%-------------------------------------------------------------

%-------------------------------------------------------------
   \begin{figure}
   \centering
\resizebox{\hsize}{!}{\includegraphics[width=15cm]{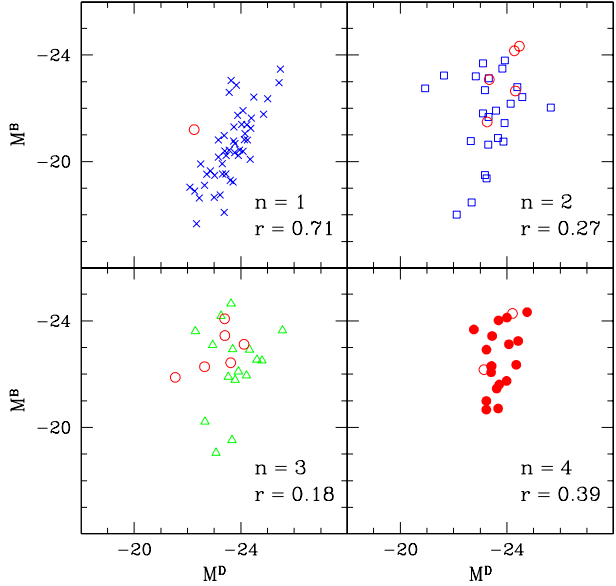}}
      \caption{Bulge versus disk absolute magnitudes $M^B$, $M^D$,
separately for galaxies with different best $n$ values of the bulge.
The correlation coefficients are shown below the values of $n$.
              }
         \label{fig:bdabs_n}
   \end{figure}
%
%-------------------------------------------------------------
The bulge $<\mu_e^B>$ significantly brighter with decreasing Hubble type;
the effect is formally highly significant\footnote{Early spiral types 
($T \leq 2$), intermediate ($2 < T \leq 4$)
and late ($4 < T \leq 6$), and very late ($T \geq 7$) types differ
at a significance level of $>99.9$\%,
with the exception of late and very late types whose difference
is significant at $\sim94$\%.},
a new result made possible by the inclusion of the MGH Sa's,
and the consequently more ample range in $T$,
as well as by our use of mean SFBs $<\mu_e^B>$, rather than 
$\mu_e$.
Although de Jong (\cite{dejong96}) found a similar trend, 
all bulges were fit with simple exponentials; it was 
therefore not clear if the effect was real for
early-type bulges which are better fit with higher $n$. 
No such trend was noted by MH01, although as they point out, 
their sample extends only to $T=5$. 
Also the disk $<\mu_e^D>$ gets brighter with decreasing Hubble type,
although the trend is much weaker and not readily discernible in 
Fig. \ref{fig:sbmean_t}\footnote{The increase in disk SFB is formally significant only for 
early ($T \leq 2$) and intermediate ($2 < T \leq 4$) types, relative
to very late types ($T\geq7$);
the first is $>99.9$\% significant, while the latter is significant 
only at a $\sim95$\% level.}.
This also is a new result, which again arises from the large range
of Hubble types in our combined sample.

Regarding both bulge $r_e^B$ and disk $r_e^D$ effective scalelengths, the variations 
with $T$ are small.
Although very late-type disks appear to be smaller than their early-type
counterparts, 
both bulge and disk scalelengths are relatively constant with Hubble type,
unlike the variation of bulge $r_e^B$ with $n$ discussed in the
previous section.
For a given Hubble type, the range spanned by disk properties
($M^D$, $<\mu_e^D>$, $r_e^D$)
is smaller than that of the bulge ($M^B$, $<\mu_e^B>$, $r_e^B$).

The lower panels of Fig. \ref{fig:sbmean_t} plot the absolute bulge and disk magnitudes
$M^B$ and $M^D$ as a function of $T$.
There is a clear trend for the bulge luminosity to decrease going from
early to very late-type spirals\footnote{The significance of the trend is $>$99.9\% 
for all bins, except
that T$\leq$2 and slightly later types are not formally distinguishable.}.
A similar result was also obtained by MH01. 
Absolute disk magnitudes do not vary significantly with T,
except that very late types T$\geq$7 are less luminous than
all earlier types\footnote{Significance is $>$99\% except for 
T$\leq$2 which has $\sim$95\% significance).}.

\subsection{Bulge parameters versus disk parameters\label{bd}} 

%Here we compare the same structural parameter for bulges and disks, 
%in order to quantify what, if any, the relation is between them, and if
%the degree of the relation has any dependence on the shape of the bulge.

Figures \ref{fig:sb_n}, \ref{fig:logre_n}, and \ref{fig:bdabs_n}
illustrate the correlations, separated by $n$, between 
the parameters of a bulge and those of its host disk.
For all parameters, there is a clear trend for $n=1$ bulges and
disks to be tightly correlated.
For $n=2$, the correlation is somewhat degraded, and by the time $n=3,4$,
the correlations have all but disappeared.
The tendency is evident for mean SFBs (Fig. \ref{fig:sb_n}), 
scalelengths (Fig. \ref{fig:logre_n}), and luminosities (Fig. \ref{fig:bdabs_n}).

While this is a new result, one of the main ones of this paper,
previous work hinted at something similar.
With 1D bulge$+$disk decompositions of R-band images,
and assigning $n=1$ or 4, Andredakis \& Sanders (\cite{andredakis94}) 
found a correlation between bulge and disk absolute magnitudes,
with the correlations being tighter for systems
with $n=1$ than for $n=4$ bulges. 
MH01 found weak correlations between $\mu_e$ and $\mu_0$, 
and with $r_e^B$ and disk scalelength $r_e^D$, but did not separate 
or bin the $n$ values; thus the trends evident in our data were diluted 
in their continuous distribution.

Correlations between bulge and disk radii (Fig. \ref{fig:logre_n}) are not new,
and have been used to argue for secular evolution
as the primary mechanism of bulge formation (e.g., Courteau et al. \cite{courteau96}).
Independently of the ratio of radii, which will
be discussed in the following section, our data show that the correlation
between radii is best for $n=1$, but degrades steadily toward higher $n$ values. 
By the time $n=4$, the correlation has disappeared completely. 
If a sample does not contain higher $n$ bulges because of a dearth of early-type spirals,  
the bulge/disk radii will appear correlated (e.g., MACH03). 
Our result also agrees with Graham (\cite{graham01}) who found a constant ratio of bulge
and disk radii only up to $n\sim2$.
Khosroshahi et al. (\cite{khosroshahi00a}) and Scodeggio et al. (\cite{scodeggio02}) also report
the existence of such a correlation although they found large scatter;
again, this is likely due to the dilution of the correlations when all
$n$ values are combined together as done by these authors.

\subsection{Bulge/disk scaling relations \label{scaling}} 

%-------------------------------------------------------------
   \begin{figure}
   \centering
\resizebox{\hsize}{!}{\includegraphics[width=9cm]{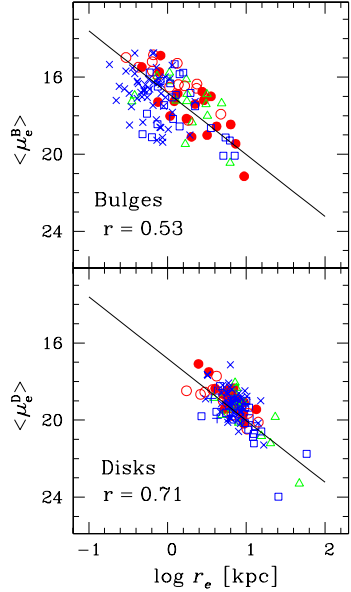}}
      \caption{Bulge mean effective surface brightness $<\mu_e^B>$
versus effective radius $r_e^B$ in kpc (upper panel).
Disk mean effective surface brightnesses $<\mu_e^D>$
versus effective radius $r_e^D$ (lower panel).
Symbols as in Fig. \ref{fig:bias}.
The best-fit regression (OLS bisector) for all data is shown
in both panels.
              }
         \label{fig:sbre_n}
   \end{figure}
%
%-------------------------------------------------------------
Figure \ref{fig:sbre_n} shows mean surface brightness $<\mu_e>$
versus effective radius, separately for bulges and disks. 
We find both bulge and disk $<\mu_e>$ to be well correlated with their effective radii, in the
sense that larger bulges and disks tend to be fainter in SFB.
This is well known; such correlations have long been found for spiral
bulges (Kent \cite{kent85}; Kodaira et al. \cite{kodaira}; Andredakis et al. \cite{andredakis95}), and 
constitute the ``Kormendy'' relation for
ellipticals (Kormendy \cite{kormendy77}; Hoessel \& Schneider \cite{hoessel}; 
Djorgovski \& Davis \cite{djorgovski}).
More recent studies (MH01; Khosroshahi et al. \cite{khosroshahi00a}) 
also find a similar correlation but
with larger scatter.
Bulges and disks appear to share the same FP, but occupy distinct, though overlapping,
regions of it (Burstein et al. \cite{burstein97}).
Combining bulges and disks together, we obtain the regression (shown in the figure) 
$<\mu_e>\,=\,16.81\,+\,3.21\,\log r_e$, that is $r_e \propto I_e^{-0.78}$,
which is consistent with canonical estimates of the FP relation
(e.g., Kormendy \& Djorgovski \cite{kd89}):
$r_e \propto I_e^{-0.83\pm0.08}$.
These are both substantially steeper than the trend reported in MH01
($r_e \propto I_e^{-1.12}$).

SFBs plotted against scalelengths form an almost face-on view of the FP
(e.g., Kormendy \& Djorgovski \cite{kd89}). 
It can be seen in Fig. \ref{fig:sbre_n} that for bulges (top panel) the main scatter in
the bulge relation is introduced by $n=1$ values ($\times$, $r=0.30$), while the best 
correlation is for $n=4$ (filled circles, $r=0.80$).
As in Carollo (\cite{carollo99}), for a given radius, $n=1$ bulges tend to be fainter
than their higher $n$ counterparts. 
As a whole, disks are significantly better correlated 
($r=0.71$) than bulges ($r=0.53$ when all $n$ are included).
Although this projection of the FP is not ideal for such considerations,
these results imply that residuals relative to the FP may be correlated with shape parameters
(e.g., Hjorth \& Madsen \cite{hjorth}; Prugniel \& Simien \cite{prugniel}; 
Khosroshahi et al. \cite{khosroshahi00b}).

%-------------------------------------------------------------
   \begin{figure}
   \centering
\resizebox{\hsize}{!}{\includegraphics[width=15cm]{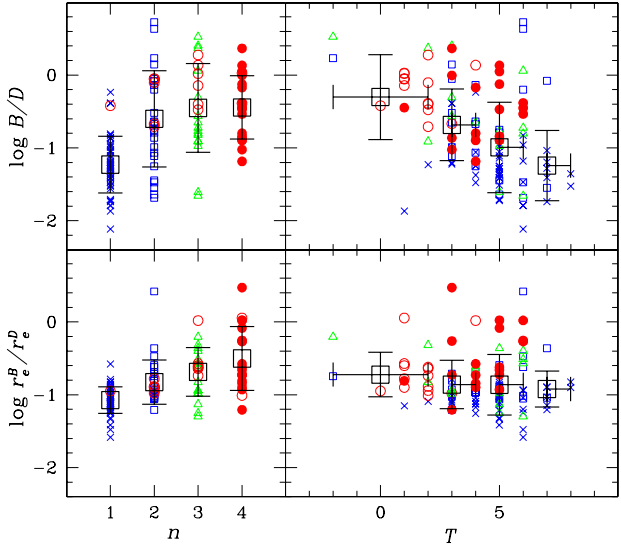}}
      \caption{Bulge and disk effective scalelength ratios $r_e^B/r_e^D$ (upper panels)
and bulge-to-disk luminosity ratios (lower) versus $n$ (left panels) and Hubble type $T$ (right).
Symbols as in Fig. \ref{fig:bias}
              }
         \label{fig:bdre_nt}
   \end{figure}
%
%-------------------------------------------------------------
Since scalelength ratios constitute the observational cornerstone for 
secular evolution scenarios (Courteau et al. \cite{courteau96}, MACH03), we shall discuss this parameter
in detail for our data set.
Figure \ref{fig:bdre_nt} shows $B/D$ and  $r_e^B/r_e^D$ as a function
of $n$ (left panels) and $T$ (right panels).
While there is a clear trend for $r_e^B/r_e^D$ to increase with $n$, 
there is only a very weak trend for smaller $r_e^B/r_e^D$ with increasing $T$.
For $n=1$ bulges, we find $<r_e^B/r_e^D>=0.085 \pm 0.037$ (51 galaxies);
for $n=2$ $<r_e^B/r_e^D>=0.15 \pm 0.10$ (31); for $n=3$ 
$<r_e^B/r_e^D>=0.21 \pm 0.16$ (18); for $n=4$ $<r_e^B/r_e^D>=0.32\pm 0.32$ (19)\footnote{
The difference in effective scalelength ratio among $n$ values is $>$99.9\% significant
for $n=1$ and larger $n$'s, and $>$99\% for $n=2$ and $n=4$.
$n=2$ and $n=3$ ratios do not differ significantly, nor do $n=3$ and $n=4$.}.
The size ratio increases from $n=1$ to $n=4$ by a factor of $\sim\,4$.
The mean value of $r_e^B/r_e^D$ we find for $n=2,3$ is 
   consistent with the MACH03 value of $0.22 \pm 0.09$ 
   for Type-I profiles, but this value is more than twice as large
as for the $n=1$ bulges in our sample. 
Independently of Type I vs. Type II profiles, we surmise that much of the scatter
in the $r_e^B$,$r_e^D$ relation is due to the mixing of different $n$ bulges.
As was seen in Section \ref{bd}, the correlation between bulge and disk scalelengths
is best for $n=1$ bulges, then gradually decays to a non-correlation for $n=4$
(see Fig. \ref{fig:logre_n}).
In addition to the trend of increasing $<r_e^B/r_e^D>$ with $n$, the {\it dispersion
also increases with $n$}:
$n=1$ bulge/disk size ratios have a scatter $\sim\,9$ times smaller than 
that for $n=4$. 

While $<r_e^B/r_e^D>$ depends on $n$, it is rather independent of Hubble type.
The means $<r_e^B/r_e^D>$ for different $T$ bins are statistically
indistinguishable, although very late-type spirals ($T\geq7$) do tend to
have $<r_e^B/r_e^D>$ smaller than very early types ($T\leq$2): 
0.12$\pm$0.07 (10 galaxies) vs. 0.19$\pm$0.13 (19 galaxies). 
MACH03 also find a slight dependence of $<r_e^B/r_e^D>$ on morphology
($\sim\,0.20-0.24$ from late- to early-type spirals), 
although the trend is very weak, and not confirmed by Graham (\cite{graham01}),
nor with formal significance by our data.
Hence, although variations of $r_e^B/r_e^D$ with $T$ are difficult to detect,
this ratio changes significantly with the shape of the bulge, implying that
the spiral sequence cannot be scale free (c.f., Courteau et al. \cite{courteau96}, MACH03).
This point will be discussed further in the context of secular evolution scenarios in
Section \ref{discussion}.

As expected, we find that $B/D$ increases with $n$ and decreases with $T$, in agreement 
with previous work (e.g., Andredakis et al. \cite{andredakis95}; 
Khosroshahi et al. \cite{khosroshahi00a}; Graham \cite{graham01}; MH01).
$B/D$ is a criterion for the Hubble classification, so the
correlation with $T$ is obvious.
The dependence of $B/D$ on $n$ is similar, although of lower significance, than its dependence 
on $T$\footnote{For our data,
the significance of the trend with $T$ is formally 99\% or greater between all type bins,
except for early type spirals ($T\leq2$) and the next bin (2$<T\leq$4), where
it is slightly lower ($\sim$97\%).
$n=1$ bulges have a significantly ($>$99.9\%) lower $B/D$ ratio than higher
$n$ values, but the differences among other $n$ values are not formally significant.}.
In any case, the spread of $B/D$ for a given $T$ bin is large, approaching two orders of magnitude,
and the looseness of $B/D$ vs. $T$ is rather surprising.
It is clear that the Hubble classification reflects complex, multivariate processes, 
which are not easily quantified. 

$B/D$ is larger for early-type spirals because of a brighter $<\mu_e^B>$, since bulge $r_e^B$,
and the disk parameters do not change significantly with $T$.
This is a confirmation of the ``iceberg'' scenario proposed by Graham (\cite{graham01}).
On the other hand, $B/D$ is larger for larger $n$ bulges because $r_e^B$ is larger,
since the disk parameters and $<\mu_e^B>$ are roughly constant.
When galaxies are divided 
according to the shape of their bulges rather than their Hubble type, 
the spiral sequence is seen to be not scale free.
It appears therefore that
the iceberg scenario proposed by Graham (\cite{graham01}) cannot be only a function of
surface brightness.
The {\it shape} of the bulge is a main factor for its increasing prominence with
decreasing $T$;
this is true not only because of the variation of $r_e^B$, but also because of the
variation in bulge central surface brightness $\mu^B_0$ due 
to the functional form of the S\'ersic law. 

%This is a confirmation of the ``iceberg'' scenario proposed by Graham (\cite{graham01}).
%, and $<\mu_e^B>$, if anything, is {\it fainter} for larger $n$. 
%Consequently, because of the variation of $r_e^B$ with $n$,
%the iceberg scenario, which is based only on surface brightness variations,
%is valid only as long as different bulge shapes are not considered.}
%Similarly to the $r_e^B/r_e^D$ ratio discussed above, 

\subsection{Luminosities \label{luminosity}}

Bearing in mind that our galaxies are rather bright objects,
with total galaxy luminosity $M^{tot}$ (H band) $\simlt -22$, 
roughly equivalent to $M^{tot}$ (B band)$\simlt\,-18$,
the sample spans roughly 4 magnitudes, providing some dynamic range for
statistical analysis.
%-------------------------------------------------------------
   \begin{figure}
   \centering
\resizebox{\hsize}{!}{\includegraphics[width=12cm]{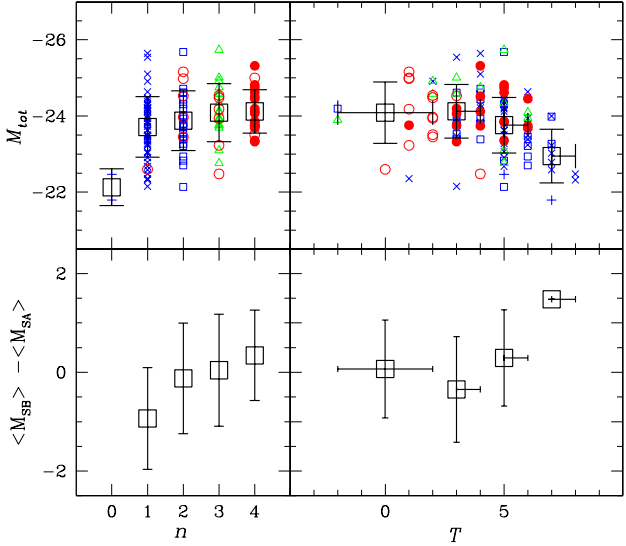}}
      \caption{Upper panels: galaxy absolute magnitudes as a function of bulge shape $n$ (left panel)
and of Hubble type $T$ (right);
symbols as in Fig. \ref{fig:bias}.
Lower panels: mean difference in absolute magnitude between strongly barred (SB) and unbarred
(SA) galaxies vs. bulge shape index $n$ (left panel) and of Hubble type $T$ (right).
              }
\label{fig:absnt}
   \end{figure}
%
%-------------------------------------------------------------
The upper panels of Fig. \ref{fig:absnt} shows the total absolute galaxy magnitude 
(sum of bulge and disk components) vs. bulge shape $n$ (left panel) and Hubble type $T$
(right).
As known for some time (e.g., Roberts \& Haynes \cite{roberts}),
there is a significant trend of luminosity with Hubble type, in that
early-type spirals tend to be more luminous than late types\footnote{In our sample, 
this trend is $>$99\% significant for all bins relative
$T \geq 7$; the difference in luminosity between
$2 < T \leq 4$ and $4 < T \leq 6$ is significant at $>95$\%, although 
$T \leq 2$ and $2 < T \leq 6$ do not have significantly different luminosities.}.
We find only a weak trend of luminosity with $n$, since
galaxies with $n=1$ bulges tend to be less luminous than higher $n$
galaxies at the $\sim95$\% level, and that the two pure disks are among 
the least luminous objects in the sample.
In any case, the range in luminosities for a given Hubble type bin is large
($\sim3$\,mag), implying that other variables are at work in addition to $T$,
e.g., DDO classification (van den Bergh \cite{vandenbergh}).

That earlier Hubble types tend to be more luminous is explained by the analogous
trend in bulge luminosity only for $T<7$, since for these types disks are of
roughly the same luminosity.
However, for $T\geq7$ disks tend to be less luminous than those of earlier types,
so that the lower total luminosities are due to a lower luminosity
of both components, not just of the bulge.

%-------------------------------------------------------------
   \begin{figure}
   \centering
\resizebox{\hsize}{!}{\includegraphics[width=8cm]{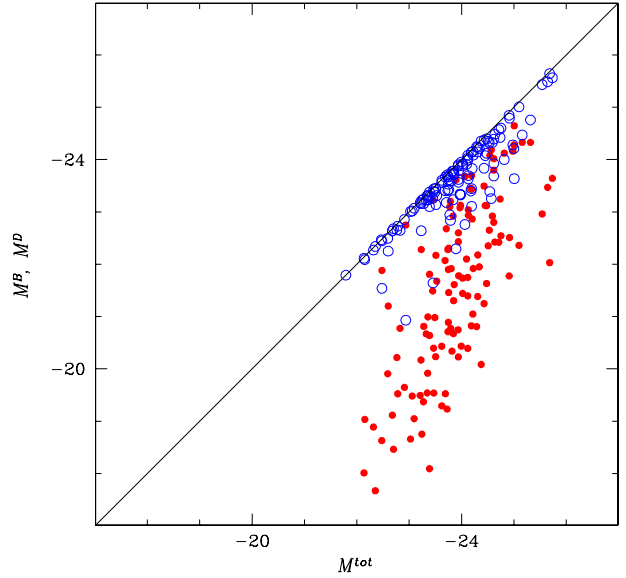}}
\caption{Bulge $M^B$ (solid circles) and disk $M^D$ (open circles)
absolute magnitudes as a function of the total 
absolute magnitude of the galaxy. Hence, there are two data points for each
galaxy. 
         \label{fig:bdmag_m}
              }
   \end{figure}
%
%-------------------------------------------------------------
$M^B$ and $M^D$ are plotted against $M^{tot}$ in Fig. \ref{fig:bdmag_m}.
The average luminosities of bulge and disk increase with the luminosity
of the parent galaxy, independently of the shape $n$ of the bulge.
The slope of the correlation is unity for disks and 
a rather steep 2.5 for bulges.
For disks associated with $n=1$ bulges, the correlation
is extremely tight since $M^D \sim M^{tot}$ for these systems. 
In addition, $M^D \sim M^{tot}$ for the less luminous disk galaxies,
while $M^D \sim M^B$ for the most luminous ones.
In other words, in the most luminous galaxies, the bulge-to-disk luminosity ratio $B/D$ $\sim$ 1.

\subsubsection{Influence of bars}

We have attempted to assess the influence, if any, of bars on galaxy luminosity,
and trends with Hubble type and bulge shape index.
In the lower panels of Fig. \ref{fig:absnt} we plot the difference between the mean
absolute luminosities $<M_{SB}>$ and $<M_{SA}>$ of the sample galaxies 
classified in RC3 as SB (strong bar) and SA (no bar), respectively;
the left panel shows this differences plotted vs. $n$, and the right one vs. $T$.
Inspection of the figure clearly shows that strongly barred galaxies 
with exponential bulges are significantly more luminous than their unbarred
counterparts.
Despite the trend of increasing luminosity with $n$ shown in Fig. \ref{fig:absnt}, 
the {\it 6 strongly barred $n=1$} galaxies are the most luminous subset in our sample:
$<M_{SB}(n=1)>\,=\,-24.5\,\pm\,0.9$.
There is also a continuous trend toward smaller differences $<M_{SB}>-<M_{SA}>$ 
with increasing $n$.
However, the same is not true for trends with Hubble type (right panel);
we find no evidence that a strong bar makes galaxies of a given type
more (or less) luminous.

\section{Discussion \label{discussion}}

First, we summarize the main results obtained in the previous section.
\begin{itemize}
\item
Bulge effective scalelengths $r_e^B$ increase with increasing $n$; $n=1$ bulges
are smaller than those with $n>1$ (Fig. \ref{fig:bd_n}).
Neither bulge and disk $<\mu_e>$ nor disk $r_e^D$ vary with $n$.
Bulge luminosity $M^B$ increases with $n$, while disk luminosity $M^D$ is independent
of $n$ (Fig. \ref{fig:bd_n}).
In particular, exponential bulges are smaller and less luminous than those with $n>1$.
\item
Bulge $<\mu_e^B>$ gets brighter as $T$ decreases (Fig. \ref{fig:sbmean_t}).
Disk $<\mu_e^D>$ shows a similar, although weaker, trend.
Bulge luminosity $M^B$ increases with decreasing $T$ (Fig. \ref{fig:sbmean_t}), 
but only disks with T$\geq$7 are significantly
less luminous than their early-type counterparts. 
Neither bulge nor disk scale lengths $r_e^B$, $r_e^D$ vary significantly with Hubble type
(Fig. \ref{fig:sbmean_t}).
\item
When bulge parameters ($<\mu_e^B>$, $r_e^B$, $M^B$) are
compared with disk ones ($<\mu_e^D>$, $r_e^D$, $M^D$), 
{\it in all cases they are tightly correlated for $n=1$ bulges} 
(Figs. \ref{fig:sb_n}, \ref{fig:logre_n}, \ref{fig:bdabs_n}).
The correlations gradually worsen with increasing $n$ such that 
$n=4$ bulges appear {\it virtually independent} of their disks.
\item
The scatter in the correlation between bulge $<\mu_e^B>$ and $r_e^B$ (Fig. \ref{fig:sbre_n})
is shown to depend on bulge shape $n$;
the two parameters are tightly correlated in $n=4$ bulges ($r=0.8$), 
and increasingly less so as $n$ decreases.
Disk $<\mu_e^D>$ and $r_e^D$ are well correlated ($r=0.7$).
\item
Luminosity ratios $B/D$ are smaller for $n=1$ bulges
than for $n>1$, and increase with decreasing $T$, but
with a surprisingly large spread (Fig. \ref{fig:bdre_nt}). 
The value of $B/D$ is governed 
by the bulge luminosity, since disks have the same average luminosity
independently of their shape index $n$.
Bulge-to-disk size ratios $r_e^B/r_e^D$ are also smaller for
$n=1$ bulges, but are independent of Hubble type. 
The value of $r_e^B/r_e^D$ for $n=1$ bulges is 4 times smaller than
that for $n=4$, with a spread which is 9 times smaller.
\item
As already known, early-type spirals tend to be more luminous than late types
(Fig. \ref{fig:absnt}); 
except for $T\geq$7, this is due to the increasing contribution of the bulge. 
\item
Strongly barred galaxies with exponential bulges are more luminous than their
unbarred counterparts. 
The difference tends to vanish with increasing $n$.
There is no analogous trend with $T$.
\end{itemize}

\subsection{Bulge formation and evolution \label{formation}}

%These results imply that exponential ($n=1$) bulges are qualitatively different than those
%with $n>1$.
%They have smaller scalelengths, lower luminosities, and their structual properties  
%are well correlated with those of their disks.
%On the other hand, their (effective mean) surface brightnesses do not differ significantly
%from those of higher $n$ bulges, and unlike $n=4$ bulges, their SFBs are not well
%correlated with their scalelengths.
%While bulges, independently of their shape $n$, are increasingly brighter
%and more luminous going from late to early Hubble types,
%their scalelengths increase substantially with $n$, independently of their Hubble type.

The better correlations of bulge and disk properties for exponential bulges suggest
that $n=1$ bulges and their disks are closely related.
The tightness of the relation worsens as $n$ increases, so that when $n=4$, bulges
appear independent of their disks.
In contrast, the (Kormendy) relation between bulge SFB and scalelength is very tight for $n=4$ bulges,
and increasingly noisy for lower $n$ values.
This implies that $n=4$ bulges are in some sense more structurally defined than those with lower $n$.

In the Introduction, we sketched the competing scenarios for bulge formation; here we
use the results of our study in an attempt to distinguish among them. 
%Since exponential bulges are smaller and less luminous than their higher $n$ counterparts,
%we infer that they are less massive.
%Such an inference is
%supported by the relatively constant mass-to-light ratio in the NIR (Gavazzi \cite{gavazzi93}).
%
In exclusively ``early'' bulge formation scenarios,
in which the bulge formed before the disk, it is difficult to understand
the close relation we observe between exponential bulges and their disks.
This relation and its {\it gradual} degradation as $n$ increases
%together with the observed differences 
%of bulge properties with $n$ ($r_e$, $M^B$) and with $T$ ($<\mu_e>$, $M^B$)
is not easily explained by either the monolithic collapse scenario or 
CDM merging of primordial disks.
If initial conditions were everything, such as in a monolithic collapse, 
bulge and disk properties should be statistically unrelated.
The same would be true
if stable disk growth depended on satellite accretion after bulge formation
from merging disks, such as in the CDM merging scenarios.
It may be that such processes operate for some (higher $n$) bulges, but they are not
a good explanation for the global properties of low $n$ ones.
Nor can they explain the gradual worsening of bulge/disk correlations with $n$.

A close relation between bulges and disks is, however, foreseen by the inside-out
formation scheme of van den Bosch (\cite{vandenbosch}).
In this scenario, the collapsing halo forms unstable disks which, through
bar instabilities, rapidly become ``bulges''.
A disk is able to form only after the bulge becomes large enough to stabilize the
ensuing bar.
Nevertheless, in this scenario, it is difficult to explain the existence of luminous
galaxies with very low $B/D$ ratios.
If the presence of a disk depends on the prior buildup of a sufficiently luminous
(massive) bulge, then we would not expect to find galaxies with luminous disks
and relatively small bulges.
Indeed, the existence of pure disk galaxies ($\sim$2\% of our sample, but see
Gavazzi et al. \cite{gavazzi00}) would appear to contradict this scenario. 
 
A more natural explanation of our observations is the secular evolution scheme in which
$n=1$ bulges are relatively ``young'' structures, having formed ``recently'' from their disks
through internal gravitational instabilities.
This process must be dissipative because of phase-space arguments
(e.g., Wyse \cite{wyse}), but may not exclusively involve bar-driven gas inflow 
(e.g., Pfenniger \& Norman \cite{pfenniger}; Combes et al. \cite{combes90}; 
Friedli \& Benz \cite{friedli93}; Norman et al. \cite{norman96}).
The evolutionary scheme proposed by Zhang in a series of papers 
(Zhang \cite{zhang96}, \cite{zhang98}, \cite{zhang99}) 
predicts that ``as a galaxy evolves through the spiral-induced collective dissipation
process, its surface density distribution (i.e., disk plus bulge) will become more and more 
centrally concentrated, together with the buildup of an extended outer envelope''.
Moreover, in a given galaxy,
the observed spread in the angular momentum distribution reflects the 
degree to which secular evolutionary processes have been at work. 
Since stable stellar systems require a greater degree of velocity anisotropy 
as $n$ decreases (Ciotti \& Lanzoni \cite{ciotti97}),
$n=1$ bulges must be kinematically less isotropic than higher $n$ ones.
If this anisotropic component is related to tangential motion,
and thus to angular momentum, such bulges must be --in the Zhang scenario-- 
{\it younger}.

Let us therefore hypothesize that exponential bulges are indeed younger 
than those of higher $n$, 
independently of the Hubble type of the galaxy in which they reside.
In other words, the first spheroids to be formed from their
disks would be bulges with $n=1$.
Such a hypothesis is supported by the blue V-H colors of $n=1$ bulges,
relative to those with $n=4$ (Carollo et al. \cite{carollo01}).
It would also be consistent with the good correlation of exponential bulge
properties with those of their disks as we have seen in Sect. \ref{bd}.
If indeed the bulge $n$ increases with time, as predicted by the Zhang models,
then our observation that $r_e^B$ and $M^B$ increase with $n$
implies that they also increase with time.
The good $n=4$ correlation of $<\mu_e^B>$ with $r_e^B$ would then be a consequence of their older age 
and better defined ``relaxed'' (virialized?) structure,
and their consequent better consistency with the FP.
The degradation of the correlations between bulge and disk parameters 
with increasing $n$ would also imply that larger-$n$ bulges are increasingly 
more distant in time from their disks.
%less well associated with (``less knowledgeable about'') their disks.
In such a scheme, the bulge and disk branches in Fig. \ref{fig:bdmag_m} could be
interpreted as the ``evolutionary tracks'' of these components.

\subsection{The Hubble sequence \label{hubble}}

Appealing as this picture may be, the trends with Hubble type still need to be explained.
Consistently with previous work, we find that higher $n$ values are more common in early-type
spirals.
Conversely, all (our sample) galaxies with $T\geq$7 have $n=1$ bulges.
Therefore, there is some indication that $n$ and Hubble type are (albeit weakly) correlated,
and may both be associated with age or evolutionary state.
Since the degree of evolution is expected to be partly determined by galaxy 
mass (Zhang \cite{zhang99}), early type
spirals which are also more luminous (massive)
should in some sense be more evolved than late types.
That {\it surface brightness increases with decreasing Hubble type for bulges and to a lesser extent
for disks} is consistent with this expectation.
For a given (constant) rate of star formation from the formation epoch to the present day,
and a sufficient gas reservoir,
we would expect higher stellar surface densities in galaxies which have been forming
stars for longer.
If, on the other hand, star-formation rates were to depend on (gas$+$stellar) mass, then we would expect
higher mass systems to have higher stellar densities.
Either way,
it is difficult to reconcile the higher stellar surface densities in early type spirals 
(Fig. \ref{fig:sbmean_t}) with an earlier evolutionary state (or younger age);
they are probably more evolved (i.e., older) systems.
Early-type spirals also tend to have lower gas fractions 
(e.g., Roberts \& Haynes \cite{roberts}) than later types, 
presumably a result of having converted the available gas into stars.

If late-type spirals ($n=1$ bulge) are younger than early types (higher $n$), 
the question of bars and their effect on evolution must be addressed, since
in local samples of galaxies, $>$75\% of very late-type spirals ($T\geq$7) are barred
(Ho et al. \cite{ho}; Hunt \& Malkan \cite{hunt99}).
%This may partly result from the instability of cool disks to bar instabilities in the absence of
%a warm spherical (stellar) component such as the bulge (Ostriker \& Peebles 1973).
The excess of bars in very late-type systems may contribute to their
morphological evolution as proposed by many groups
(e.g., Combes et al. \cite{combes90}; Pfenniger \& Friedli \cite{friedli91}; 
Friedli \& Benz \cite{friedli93}; 
Hasan et al. \cite{hasan}; Norman et al. \cite{norman96}).
%Numerous numerical simulations suggest that as bars evolve, they tend to ``dissolve''
%either into lenses (e.g., Kormendy 1982) or triaxial structures which resemble bulges.
Numerical models show that bulge-like structures with an exponential $n=1$ profile can be
produced from bar-like perturbations that induce inward radial inflow of disk matter
(Pfenniger \& Friedli \cite{friedli91}; Zhang \& Wyse \cite{zhangwyse}).
Our finding that strongly barred $n=1$ bulges tend to be more luminous than unbarred ones
could reflect such evolution.

Indeed, this bar-driven evolutionary scheme has been advocated by 
Courteau et al. (\cite{courteau96}) and MACH03 to explain 
the relative constancy of $B/D$ size ratios they find in 
(the Type I profiles of) their late-type spiral sample.
However, they claim that the ratio $r_e^B/r_e^D$ is constant for all spiral types, with the implication that
the Hubble sequence is scale free.
One of the main results of our study is that several of the trends observed are
clearly detected as a function of bulge shape $n$, rather than $T$; we interpret
this as due to the complex interplay of the spiral classification criteria.
With a sample containing few early-type ($T<2$) spirals, and thus few high-$n$ bulges, several
of the trends we found would not be detectable.
Hence, we find a wider range of $n$, in agreement with other groups
(Andredakis et al. \cite{andredakis95}; MH01; Graham \cite{graham01}), and also a wider range of $r_e^B/r_e^D$.
We also find that the bulge-to-disk size ratio changes significantly with $n$.
While such variations with $n$ indicate that the Hubble sequence is not
scale free, they do not necessarily disprove the secular evolution scenario.
However, the scenario must be expanded to include the genesis of
bulges with different (higher) $n$'s, together with their distinct properties.

Bar-driven inflow as the unique mechanism of secular evolution may not be sufficient
to explain the full range of properties in the Hubble sequence of spirals.
The analytical models of Zhang \& Wyse (\cite{zhangwyse}) based on viscous evolution of bars find no galaxies 
with $B/D \sim 1$; their bulges are not sufficiently luminous compared to their disks.
Moreover, the secular-evolution simulations of Bouwens et al. (\cite{bouwens}) which rely on 
similar assumptions also produce bulges which are too small and too blue. 
The collective stellar dissipation scheme proposed by Zhang (\cite{zhang99}) may also be
insufficient. 
While qualitatively consistent with our findings, this scheme
produces effects (e.g., buildup of central concentration)
which are much smaller than those we observe, perhaps because the effects of gas are not included.

Evidently, some other mechanism is needed, such as mergers 
(e.g., Baugh et al. \cite{baugh96}) or accretion episodes (e.g., Steinmetz \& Navarro \cite{steinmetz}).
However, even mergers in hierarchical clustering scenarios are not able to generate the 
range of bulge/disk properties in our sample.
Scannapieco \& Tissera (\cite{scannapieco03}) use hydrodynamical simulations to 
assess the effects of mergers on
disk galaxy structure.
While they are able to reproduce the bulge-to-disk size relation for the $n\leq2$ bulges in our sample and in 
that of MACH03, the tight correlation of $r_e^B/r_e^D$ for $n=1$ bulges is not explained. 
Moreover, they fail to predict the strong increase of $r_e^B$ with $n$ 
(Fig. \ref{fig:bd_n} and other work), since their
models predict the opposite trend.
Nevertheless, they successfully reproduce the $B/D$ ratio as a function of bulge shape, although 
their $n$ values tend to cluster around 1, unlike the early-type spirals in our sample which tend to be best
fit by larger $n$ values (Fig. \ref{fig:T_n}).
On the other hand Aguerri et al. (\cite{aguerri01}) are able to reproduce $n=4$ bulges with satellite
accretion, but only with satellites as massive as the bulge itself.

\subsection{Conclusions \label{conclus}}

The hypothesis that emerges from our study is essentially the following:
\begin{itemize}
\item
Disks are formed first, during an initial collapse, and have properties which
depend on the mass and angular momentum of the dark-matter (DM) halo.
\item
Bulges are then born first as $n=1$ systems, through secular
transfer of disk material to the central regions, probably by bar instabilities.
\item
With an evolutionary rate that depends on initial conditions (gas+DM mass, angular momentum),
$n$ of the bulge increases, with consequent growth of $r_e^B$, $M^B$, and $B/D$,
perhaps by means of the collective dissipation mechanism described by Zhang (\cite{zhang99}),
together with the effects of gas transfer.
\item
As $n$ gets larger and the bulge builds up, Hubble type gets earlier, with consequent 
brightening (SFB) of bulges and (less so) disks, and an increase of $M^B$, $B/D$.
\item
Since $n=1$ bulges are relatively unevolved systems, in a state of formation at the
present epoch, they are
structurally not as relaxed as larger $n$ bulges, and present a large scatter in
the $<\mu_e^B>-\log\,r_e^B$ (Kormendy) relation.
Their stellar luminosity is enhanced by the presence of a strong bar 
driving the evolution.
The $n=1$ bulges that we see today are those systems which have had long evolutionary
times, probably because of their lower luminosity (mass).
The evolutionary state of the disk would also be a function of $n$, since 
the entire system would be expected to have evolved faster when $n$ is larger.
\item
Mergers of bulge/disk systems and accretion of gas probably affect the buildup of the
bulge, but are probably important only for the creation of high-$n$ bulges.
\item
The correlation between ``evolutionary state indicators'' $n$ and $T$
is not expected to be perfect because of the complex nature of spiral classification, 
and because of
variations in galaxy mass, angular momentum, bar strength, gas fraction, dark halo merging 
history, and perhaps other parameters.
\item
The gradual growth of a bulge as $n$ increases is implied by the
continuity we observe in bulge properties as a function of $n$.
All trends are tight for $n=1$, and steadily worsen with increasing $n$, 
except for the Kormendy relation which is best followed by structurally 
evolved systems ($n=4$).
\end{itemize}

One prediction of such a scenario is a certain spread in 
stellar age for high-$n$ bulges.
Detailed simulations are necessary to assess whether this prediction is
in conflict with  
%Rather there is strong evidence (e.g., the color-$\sigma$ relation)
the level of synchronicity in the star-formation history observed in
spheroids (Bower et al. 1992).
We also would expect that exponential bulges, prevalent in late-type spirals 
and closely connected with their disks, 
would be more dominated by rotation than their higher $n$ counterparts.
Low-luminosity spheroids are known to have a higher degree of rotational
support than more luminous ones (Davies et al. 1983),
so it may
be that this is true also for $n=1$ bulges relative to those of higher $n$. 

Clearly more modelling is needed to understand the full range in properties along the Hubble
sequence, and what they imply for bulge formation.
However, secular evolution appears to be a strong driver of the Hubble sequence. 
Our observations are compatible with the hypothesis that $n$ and $T$ are both rough measures of bulge age,
and that bulges are, at least initially ($n=1$), formed secularly from disks. 

\begin{acknowledgements}
      We warmly thank G. Moriondo who compiled this sample for his Ph.D. thesis,
	and performed many of the observations, all the data reduction, and
	much of the preliminary analysis. 
We also thank the ARNICA team for making the camera possible, and the various telescope
committees for generous time allocations.
\end{acknowledgements}

\begin{figure*}
{\rotatebox{180}{\resizebox{17.5cm}{!}{\includegraphics*[30,80][530,770]{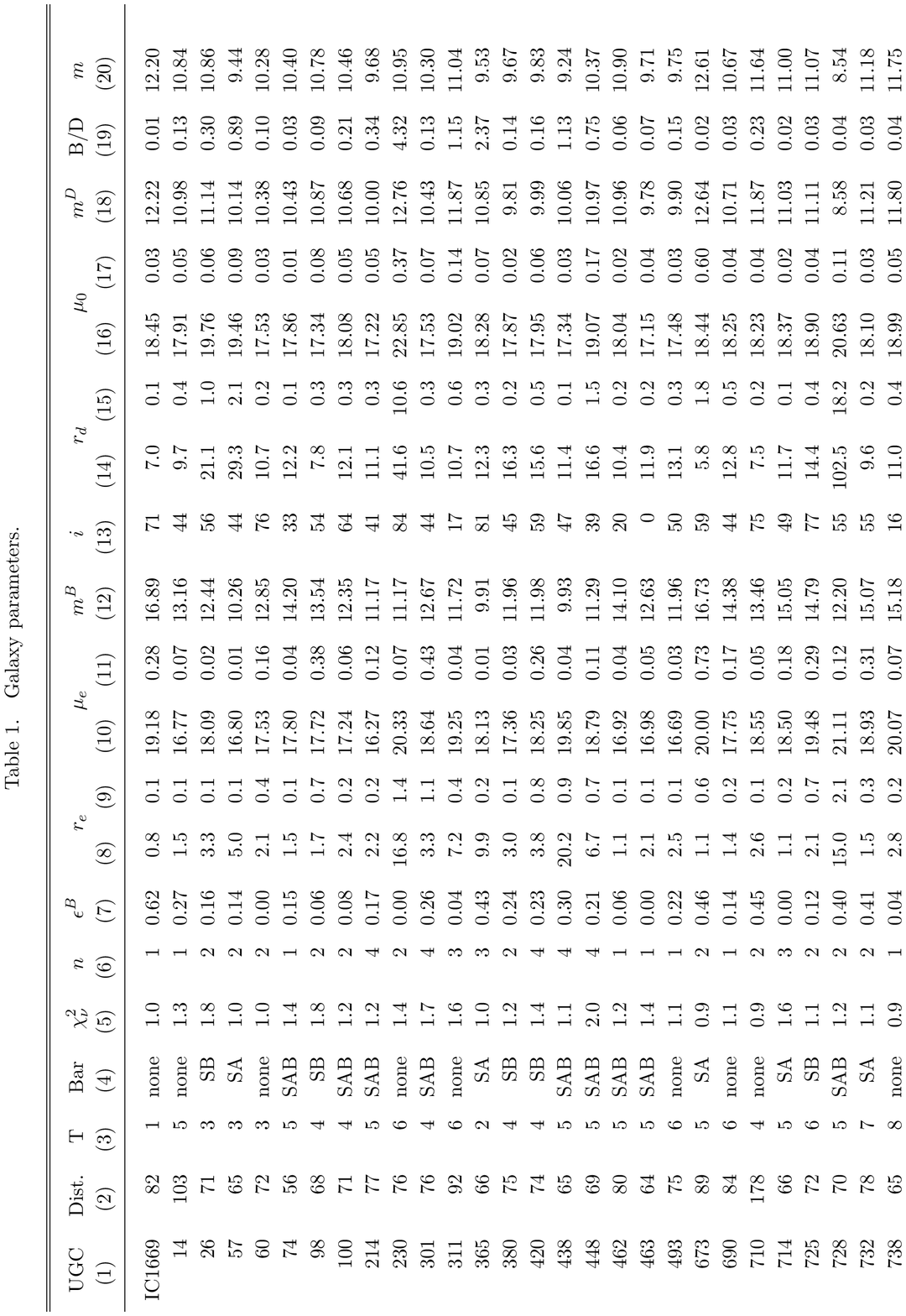}}}}
\end{figure*}
\begin{figure*}
{\rotatebox{180}{\resizebox{17.5cm}{!}{\includegraphics*[40,80][540,770]{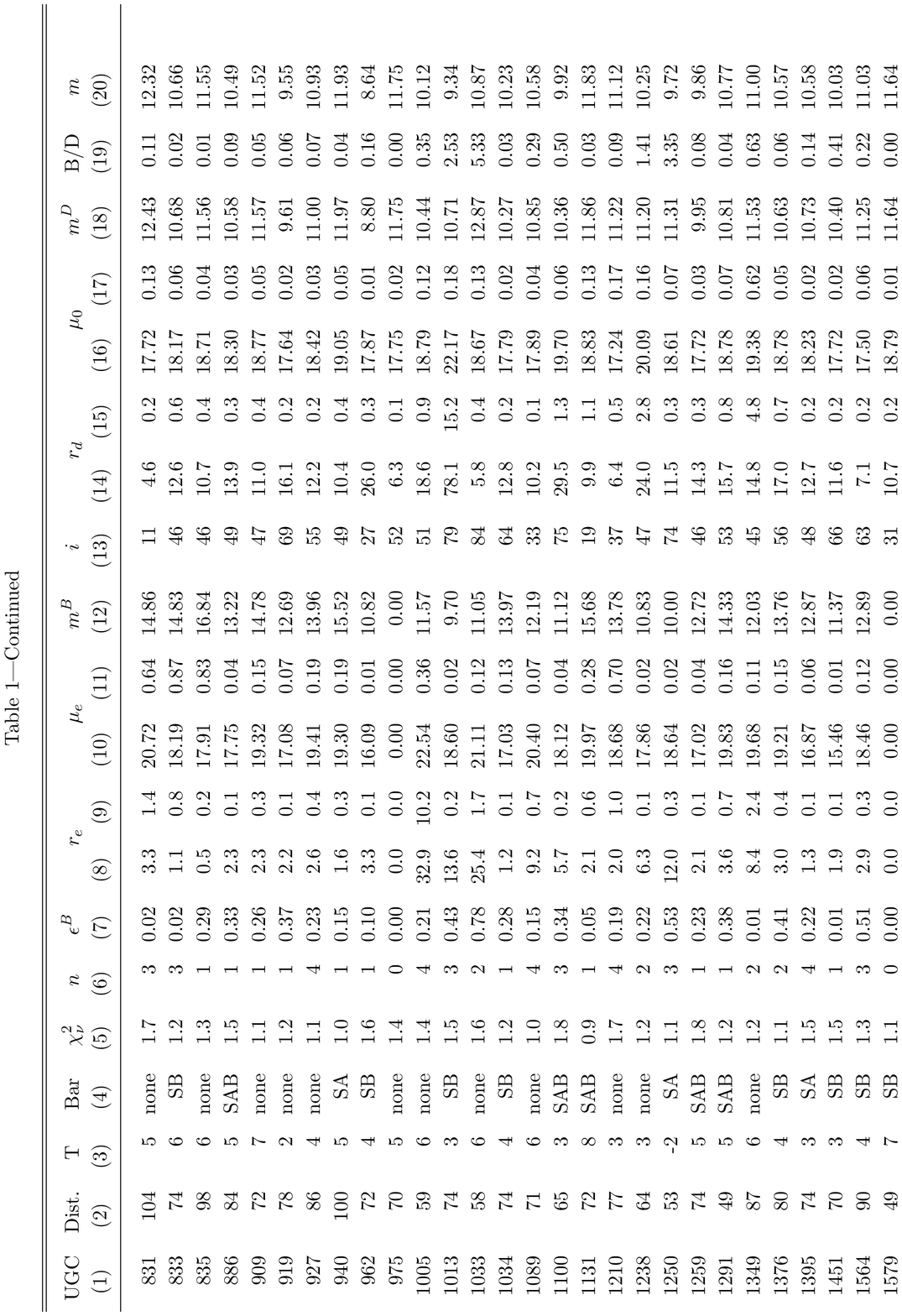}}}}
\end{figure*}
\begin{figure*}
{\rotatebox{180}{\resizebox{17.5cm}{!}{\includegraphics*[40,80][540,770]{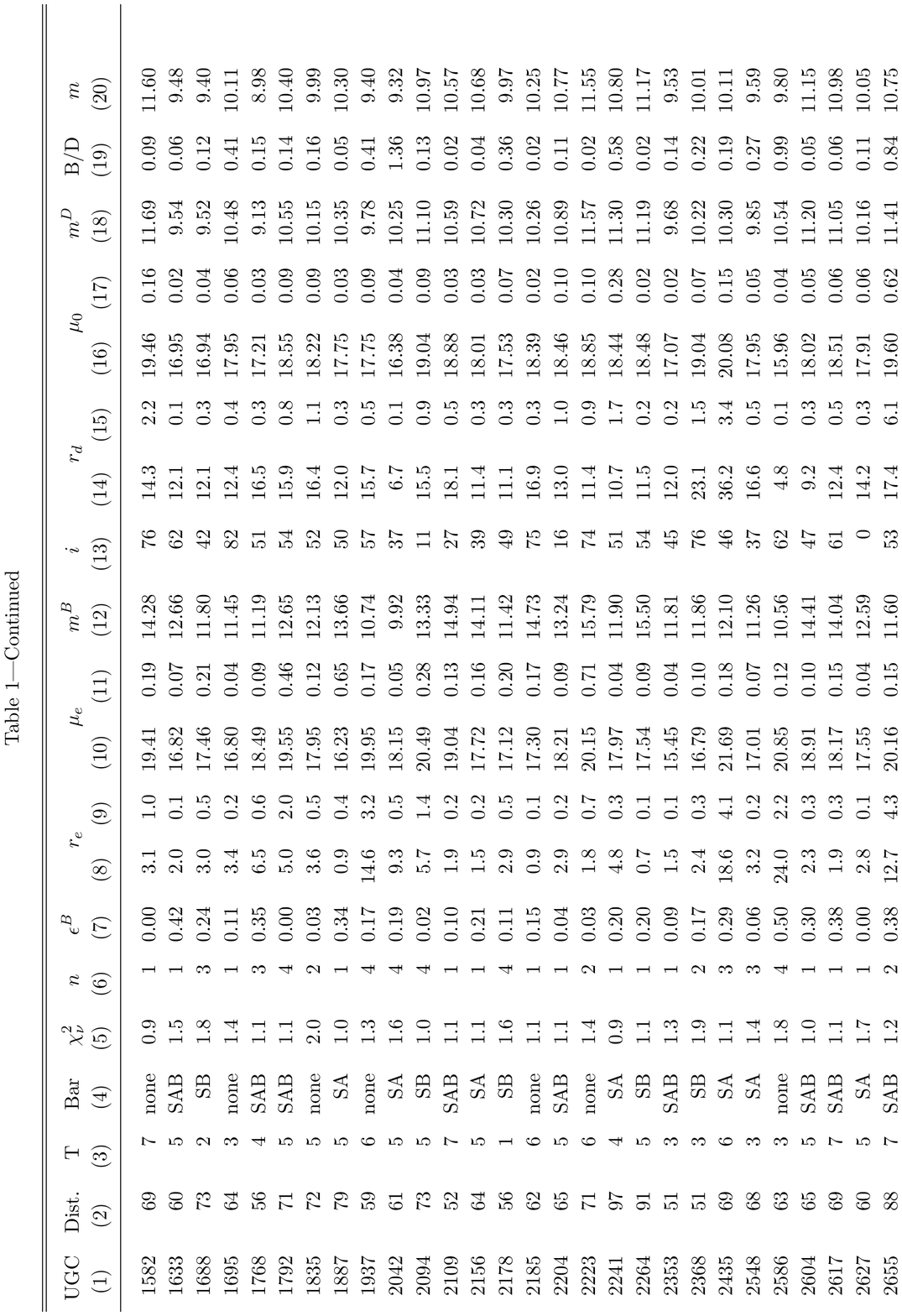}}}}
\end{figure*}
\begin{figure*}
{\rotatebox{180}{\resizebox{17.5cm}{!}{\includegraphics*[40,90][540,770]{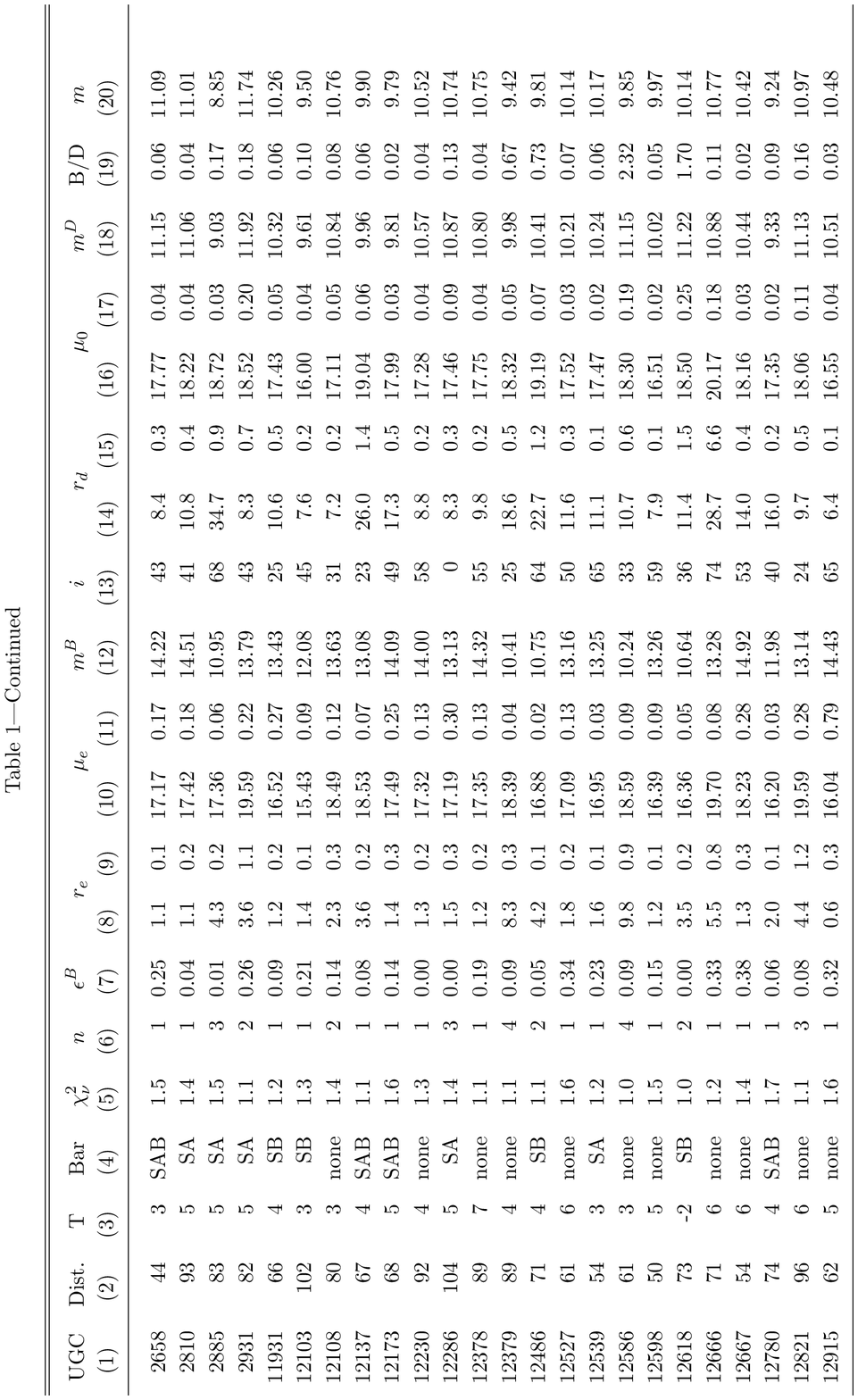}}}}
\end{figure*}

\end{document}